\documentclass[aps,prb,twocolumn,notitlepage,superscriptaddress,showpacs]{revtex4-1}
\usepackage{amsmath,latexsym,amssymb,bm}
\usepackage{hyperref}
\usepackage{color}
\usepackage{graphicx,subfigure}
\usepackage{multirow}
\usepackage{enumerate}

\usepackage{natbib}

\begin{document}

\title{Two-hole ground state wavefunction: Non-BCS pairing in a $t$-$J$
two-leg ladder}
\author{Shuai Chen}
\affiliation{Institute for Advanced Study, Tsinghua University, Beijing,
100084, China}
\author{Zheng Zhu}
\affiliation{Department of Physics, Massachusetts Institute of Technology,
Cambridge, Massachusetts, 02139, USA}
\author{Zheng-Yu Weng}
\affiliation{Institute for Advanced Study, Tsinghua University, Beijing,
100084, China} 
\affiliation{Collaborative Innovation Center of Quantum
Matter, Tsinghua University, Beijing, 100084, China}
\date{\today }

\begin{abstract}
Superconductivity is usually described in the framework of the Bardeen-Cooper-Schrieffer (BCS)
wavefunction, which even includes the resonating-valence-bond (RVB)
wavefunction proposed for the high-temperature superconductivity in the cuprate.
A natural question is 
\emph{if} any fundamental physics could be possibly missed by applying such
a scheme to strongly correlated systems. Here we study the pairing
wavefunction of two holes injected into a Mott
insulator/antiferromagnet in a two-leg ladder using variational Monte Carlo
(VMC) approach. By comparing with density matrix renormalization
group (DMRG) calculation, we show that a conventional BCS or RVB pairing of
the doped holes makes qualitatively wrong predictions and is incompatible with the fundamental pairing force
in the $t$-$J$ model, which is kinetic-energy-driven by nature. By
contrast, a non-BCS-like wavefunction incorporating such novel effect will result
in a substantially enhanced pairing strength and improved ground state
energy as compared to the DMRG results. We argue that the non-BCS form of such a new ground state
wavefunction is essential to describe a doped Mott
antiferromagnet at finite doping.

\end{abstract}

\maketitle

\section{Introduction}

Three decades after the discovery of high-temperature superconductivity in
the copper oxide materials \cite{Bednorz1986}, it still remains a mystery  whether the superconductivity can be described by a
wavefunction of Bardeen-Cooper-Schrieffer (BCS) type\cite{BCS}. For example, as a
non-phonon mechanism, the resonating-valence-bonds (RVB) ground state
proposed by Anderson \cite{Anderson1987} may be still regarded as the
BCS-like, only subject to a Gutzwiller projection onto a restricted Hilbert
space to enforce the no double occupancy of the electrons. Such a projection is due
to the on-site Coulomb repulsion $U$, which will make the electrons form
an insulating antiferromagnetic state (Mott insulator \cite{Mott1949basis,Imada1998metal}) at half-filling, where the condensate of the Cooper pairs
reduces to that of the neutral spin RVB pairing. The true Cooper pairing
similar to a conventional superconductor is expected \cite{Anderson1987} to emerge by charging the neutral RVB
background upon doping. Such an ``RVB pairing mechanism'' of
superconductivity has been intensively studied\citep{Fradkin1990short, Wen2006doping} based on the variational RVB state \cite{Anderson1987, Sorella2002superconductivity,Anderson2004physics, edegger2007gutzwiller,Scalapino2012common}.

Taking an instructive limit of two holes injected into the half-filled spin
background, one may examine the RVB origin of pairing by the following
variational construction: 
\begin{equation}  \label{rvbgs}
|\Psi_{\mathrm{BCS}}\rangle_{2\text{h}}=\hat {\Delta} |\text{RVB}\rangle,
\end{equation}
in which the two doped holes form a Cooper pair 
\begin{equation}  \label{delta}
\hat {\Delta}=\sum_{i,j} g(i,j) c_{i\uparrow}c_{j\downarrow}~,
\end{equation}
on a \emph{half-filling} insulating ground state denoted by $|\text{%
RVB}\rangle$. Here $|\text{RVB}\rangle$ is governed by the Heisenberg superexchange term with the
coupling constant $J$, which is assumed \cite{Anderson1987} to be the main driving
force for the Cooper pairing of doped holes. Namely, the antiferromagnetic correlations in $|\text{RVB}\rangle$ can provide a \emph{bare} binding force for the two holes injected into such a spin background.
Then Eq.~(\ref{rvbgs}) may serve as an important test of the RVB mechanism. To this end, the pair amplitude $%
g(i,j) $ is taken as a c-number, which can be determined variationally by using the
variational Monte Carlo (VMC) method \cite{Anderson2004physics, Sorella2005wave, edegger2007gutzwiller,Scalapino2012common} based on the $t$-$J$ model description
of the doped Mot insulator. 

However, the ansatz state in Eq.~(\ref{rvbgs}) does not necessarily
capture the fundamental physics of two hole pairing\citep{Weng2011superconducting, Weng2011mott, Zaanen2011mottness}. The key
assumption there is that the quantum fluctuation is negligible
such that $g(i,j)$ may be simulated by a ``mean-field'' in the variational approach.
However, a recent density matrix renormalization group (DMRG) study on the
ground state of two holes \citep{Zhu2014pairing,Zhu2017pairing} has revealed a different
nature of pairing other than Eq.~(\ref{rvbgs}). For such a strongly
correlated model in which two holes are injected into two distinct Mott
insulators of two-leg ladder systems, a strong \emph{phase fluctuation} has
been identified \citep{Zhu2017pairing} in the pair-pair correlation
functions. It suggests \citep{Zhu2017pairing}  that the pair amplitude $g(i,j)$ in Eq. (\ref%
{delta}) should be replaced by 
\begin{equation}  \label{g}
g(i,j)\rightarrow g(i,j)e^{-i\left(\hat{\Omega}_i+\hat{\Omega}_j\right)}~,
\end{equation}
where $\hat{\Omega}_i$ represents a nonlocal phase shift produced by doping
a hole into the system. Here $\hat{\Omega}_i$ has been explicitly identified \citep{Zhu2017pairing}
as a pure spin string operator [cf.~Eq.~(\ref{PS})] acting on the
half-filling background $|\text{RVB}\rangle$, and is very sensitive to the
spin-spin correlation in $|\text{RVB}\rangle$. In essence, it implies that
the correct two-hole ground state should be properly characterized by
\begin{equation}  \label{gs}
|\Psi_{\mathrm{G}}\rangle_{2\text{h}}=\hat {D} |\text{RVB}\rangle~,
\end{equation}
where 
\begin{align}\label{D}
\hat {D} =\sum_{i,j} {g}(i,j) \tilde{c}_{i\uparrow}\tilde{c}_{j\downarrow}~,
\end{align}
is equal to $\hat {\Delta}$ in Eq.~(\ref{delta}) with 
\begin{equation}\label{ctilde}
{c}_{i\sigma}\rightarrow \tilde{c}_{i\sigma}\propto {c}_{i\sigma}e^{-i\hat{%
\Omega}_i} ~.
\end{equation}
Namely, the Cooper pairing of two bare holes in the BCS-like ground state 
(\ref{rvbgs}) should be replaced by the pairing of two new ``twisted''
quasiparticles, created by $\tilde{c}_{i\uparrow}$ and $\tilde{c}% 
_{i\downarrow}$ on the ``vacuum'' $|\text{RVB}\rangle$. In other words, each doped hole has to change the spin background $|\text{RVB}\rangle$ by a nonlocal phase shift  $\hat{\Omega}_i$ to become a true quasiparticle. Due to the spin-dependent many-body phase shift operator $e^{-i\hat{\Omega}_i}$, which is non-perturbative by nature\citep{Zhu2017pairing}, the resulting new ground state (\ref{gs}) is obviously non-BCS-like in the original electron representation.

Similar novel quantum phase fluctuations have been also identified in a symmetry-protected topological phase of the two-leg system \citep{Zhu2017pairing}, in which two spins at each rung are coupled by ferromagnetic instead of antiferromagnetic coupling. It implies that the pairing structure may be generally of non-BCS-type in a doped spin system enforced by the no-double-occupancy constraint. Recently, the pairing of holes at finite doping has been clearly found  by DMRG in various generalized doped Mott insulators\citep{White1997ground, Patel2017pairing,  Jiang2017, tohyama2018dynamical, Jiang2018Fourleg}. It is thus highly intriguing and motivating to understand the microscopic origin of the hole pairing state in the limit of a two-hole case, which should shed light on the superconducting mechanism and the wavefunction structure at finite doping, which are experimentally relevant. 

%------------------------------------------------------
\begin{figure}[t]
\begin{center}
\includegraphics[width=0.5\textwidth]{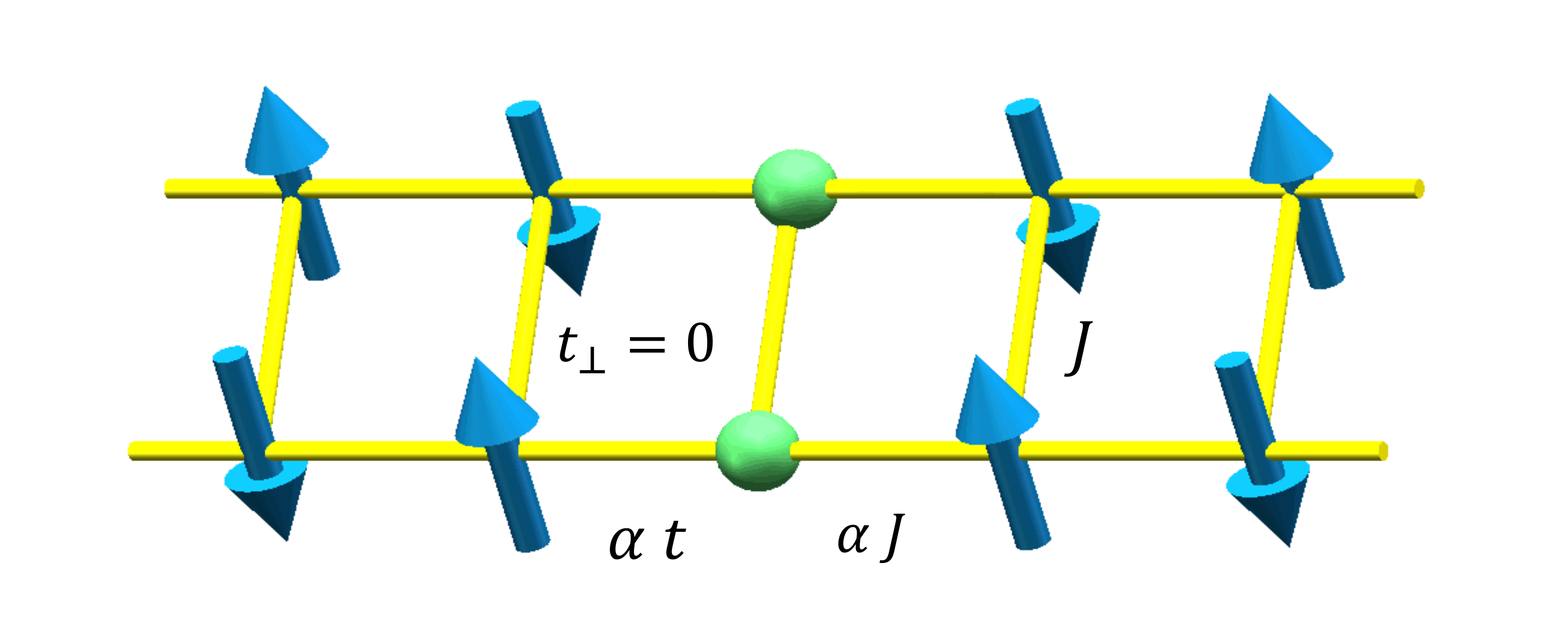}
\end{center}
\par
\renewcommand{\figurename}{Fig.}
\caption{(Color online.) The structure and parameters for a two-leg $t$-$J$
square ladder doped by two holes. Here, the intrachain hopping and superexchange coupling constants are denoted by $\protect\alpha t$ and $\protect\alpha J$, respectively, with $\alpha>0$ as the anisotropic parameter and the interchain superexchange coupling defined by $J$. Note that the interchain hopping $t_{\perp }=0$ in the present work. }
\label{Fig1}
\end{figure}
%--------------------------------------------------------

In this paper, we comparably study the two variational ground states, Eq.~(%
\ref{rvbgs}) and Eq.~(\ref{gs}), by the VMC approach
based on the $t$-$J$ model. Specifically, the half-filling ground state $|%
\text{RVB}\rangle $ will be first determined in a Heisenberg two-leg square
ladder model as illustrated in Fig.~{\ref{Fig1}}, which describes a short-range antiferromagnetic or RVB spin state. 
Then we examine the two-hole ground state with turning on the hopping integral along the
chain direction (but without the inter-leg hopping along the rung direction
for the simplicity in analytic analysis). We variationally determine the
parameter $g(i,j)$ by minimizing the two-hole ground state energies of Eq. (\ref{rvbgs}) and Eq. (\ref{gs}), respectively. We find that the ground state energy and various pair-pair correlations of the ground state (\ref{gs})
are significantly and qualitatively improved over the BCS pairing state (\ref {rvbgs}), in excellent agreement with the DMRG results. In particular, by using a unitary transformation, we show that
the ground state (\ref{gs}) properly incorporates the
kinetic-energy-driven pairing force hidden in the $t$-$J$-type model, which
is completely missed in the RVB-like description in Eq.~(\ref{rvbgs}). In fact, in the latter state, we show variationally that two holes do not form a bound state at all, even though $|\text{RVB}\rangle$ as an RVB state possesses
the same short-range antiferromagnetic correlation in the two-leg ladder.  In other words, a new pairing mechanism distinct from the RVB mechanism can be explicitly identified in the strong binding state of Eq.~(\ref{gs}), which is argued to be Amperean-like  \cite{Lee2014amperean,Lee2007amperean}.  Generalizations to the $t_{\perp}\neq 0$ or the two-dimensional case, as well as the finite doping case, will be also briefly discussed.  

The rest of the paper is organized as follows. In Sec. \ref{Sec::model} A, we
first introduce a $t$-$J$ type model for the two-leg ladder illustrated in Fig. {\ref{Fig1}} and the corresponding $\sigma\cdot t$-$J$ model for the purpose of comparison. Then, in Sec. II B, we study the ground state properties of two different types of variational wavefunction, $|\Psi_{\mathrm{BCS}}\rangle_{\mathrm{2h}}$ and $|\Psi_{\mathrm{G}}\rangle_{\mathrm{2h}}$ outlined in the Introduction, by the VMC calculation. By making a comparison with the DMRG results, we show that the latter ground state does capture the essential physics especially the non-BCS pairing in the $t$-$J$ type model, whose nature is further analyzed in Sec. II C. Finally,  the summary and discussion of the main results, as well as some perspectives, are given in Sec. III.

\section{Model and Results\label{Sec::model}}

\subsection{The model}

In this paper, we mainly focus on the ground state properties of the two-hole-doped Mott insulator on a
two-leg-square-ladder illustrated in Fig. {\ref{Fig1}}, which is described by the $t$-$J$ Hamiltonian\citep{Zhu2015quasiparticle, Zhu2017pairing} $H_{t\text{-}J}=H_t+H_J$ as follows:
\begin{align}  \label{t-J}
H_{t} & =-\alpha t\sum_{i\sigma}\left( c_{1i\sigma}^{\dagger}c_{1i+1\sigma
}^{}+c_{2i\sigma}^{\dagger}c_{2i+1\sigma}^{}+\text{H.c}\right)\\
H_{J} & =\alpha J\sum_{i}\left( \mathbf{S}_{1i}\cdot \mathbf{S}_{1i+1}+%
\mathbf{S}_{2i}\cdot \mathbf{S}_{2i+1}\right) +J\sum_{i}\mathbf{S}_{1i}\cdot 
\mathbf{S}_{2i} 
\end{align}%
where the subscripts, $1$ and $2$, label the two legs and the anisotropic
parameter $\alpha>0$ can continuously tune the spin-spin correlation length along the chain direction in the quantum spin background. (Note that if one starts with a large-$U$ Hubbard model with an $\alpha$-dependent hopping, $\alpha$ in $H_J$ should be replaced by $\alpha^2$ instead.  Previous investigations \cite{Zhu2015charge} have shown that the two models are quantitatively similar provided that $\alpha$ is not much larger than 1.)
$\mathbf{S}_{i}$ denotes the spin operator and $c_{i\sigma}$ is the hole creation operator
at site $i$ with spin index $\sigma$. The Hilbert space should satisfy the no-double-occupancy
constraint $\sum_{\sigma}c_{1,2i\sigma}^{\dagger}c_{1,2i\sigma}\leq1$. We choose the typical ratio $t/J=3$, while, for simplicity, the injected holes are only allowed to move along the chain (leg) direction with the rung hopping integral $t_{\perp}=0$ (cf. Fig. {\ref{Fig1}}). 

Previously the corresponding two-hole ground state has been studied numerically by DMRG in Ref.~\onlinecite{Zhu2017pairing} for $t_{\perp}=0$, and in Ref.~\onlinecite{Zhu2014pairing} for the general case at $t_{\perp}=t$, respectively. In both cases, a strong binding between the two injected holes has been well established by DMRG \cite{Zhu2014pairing,Zhu2017pairing}. By contrast, in these numerical studies, it has been shown that the pairing between the holes will get substantially weakened \cite{Zhu2017pairing} or even disappear \cite{Zhu2014pairing} if the hidden phase-string sign structure 
in the $t$-$J$ model is precisely removed in the so-called $\sigma\cdot t$-$J$ model defined by $H_{\sigma\cdot t\text{-}J}=H_{\sigma \cdot t}+H_J$, in which the superexchange Hamiltonian $H_J$ remains the same, but the hopping term is changed to  \cite{Zhu2014pairing,Zhu2017pairing}
\begin{equation}  \label{st}
H_{\sigma \cdot t}=-\alpha t \sum_{\sigma i}\sigma \left(c_{1i\sigma
}^{\dagger }c_{1i+1\sigma }^{{}}+c_{2i\sigma }^{\dagger}c_{2i+1\sigma }^{{}}+%
\text{h.c.}\right)~
\end{equation}
by inserting a spin-dependent sign factor $\sigma=\pm 1$ in the original
hopping term of Eq. (\ref{t-J}). Then the novel non-BCS-pairing mechanism hidden in the $t$-$J$ model will lie in the  distinction between the $t$-$J$ and $\sigma\cdot t$-$J$ model, which can be effectively revealed by using the $\sigma\cdot t$-$J$ model as a useful reference Hamiltonian in the following variational study.

%%%%%%%%%%%%%%%--FIG--%%%%%%%%%%%%%%%%%%%%%%%%%%%%
\begin{figure}[tbp]
\includegraphics[width=0.5\textwidth]{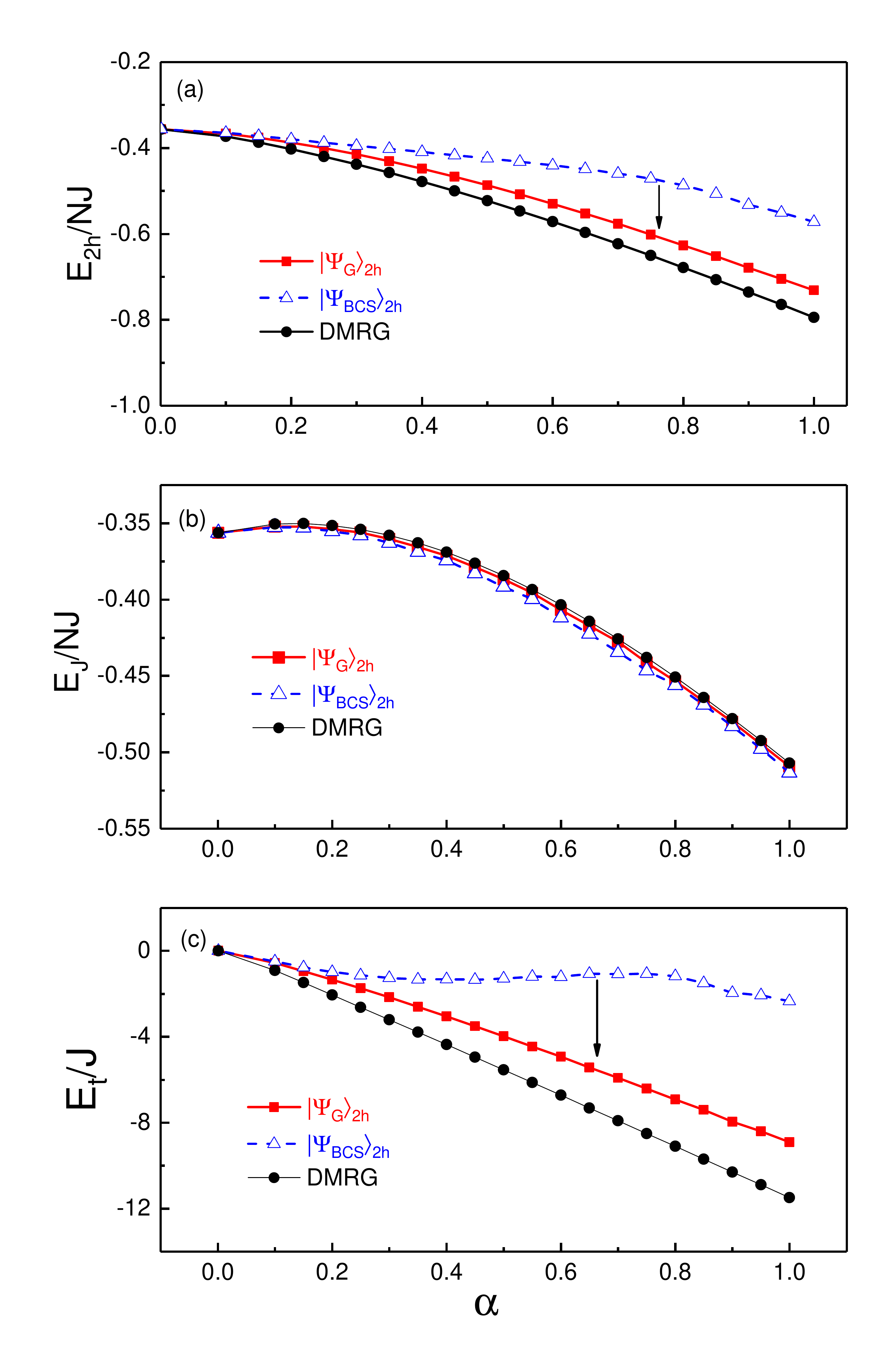}
\par
\renewcommand{\figurename}{Fig.}
\caption{(Color online.) The variational ground state energies of  $|\Psi_{\mathrm{BCS}%
}\rangle_{\mathrm{2h}}$ in Eq. (\ref{rvbgs}) (blue triangle) and $|\Psi_{\mathrm{G}}\rangle_{\mathrm{2h}}$ in Eq. (\ref{gs}) (red square) in comparison with the DMRG results (solid circle). (a)
The total energies, (b) the superexchange energies, and (c) the kinetic energies. The ladder size is $N=20\times 2$ under the open boundary condition.}
\label{Energy}
\end{figure}
%%%%%%%%%%%%%%%%%%%%%%%%%%%%%%%%%%%%%%%%%%%%%%

\subsection{Ground state wavefunctions:  Variational Monte Carlo calculation}

As pointed out in the Introduction, the ground state (\ref{rvbgs}) depicts
the simplest pairing wavefunction of two holes doped into an RVB
(short-ranged antiferromagnetic) background as envisaged originally by
Anderson \cite{Anderson1987}. By contrast, the ground state (\ref{gs}) is modified non-perturbatively by that each
doped hole induces a nonlocal phase shift as given by \citep{Zhu2017pairing} 
\begin{equation}
\tilde{c}_{\gamma i\sigma }=c_{\gamma i\sigma }e^{-i\hat{\Omega}_{\gamma i}}
\label{tildc}
\end{equation}%
and 
\begin{equation}
\hat{\Omega}_{\gamma i}=\pi \sum_{l>i}n_{\gamma l}^{\downarrow }~,
\label{PS}
\end{equation}%
where the subscript $\gamma =1,2$ labels the two legs of the square ladder
shown in Fig. {\ref{Fig1}}, and $n_{\gamma l}^{\downarrow }$ denotes the
number operator of a down spin at site $l$ along the chain of leg $%
\gamma $ \citep{Zhu2017pairing}. Note that $\hat{\Omega}_{\gamma i}$ is
taken as purely one-dimensional here in Eq.~(\ref{PS}) simply
because the hopping integral $t_{\perp }=0$ along each rung of the ladder
(cf. Fig. {\ref{Fig1}}). In general with $t_{\perp }\neq 0$, the spins in
another chain of the two-leg ladder will also play a non-negligible role in $\hat{%
\Omega}_{\gamma i}$ in general \cite{wang2015variational}. 

Note that at half-filling, both the ground states of Eqs.~(\ref{rvbgs}) and~(\ref{gs})
reduce to the same $|\text{RVB}\rangle $, which can be accurately determined
based on the Liang-Doucot-Anderson bosonic RVB wavefunction\citep{Liang1988some} for the two-leg
Heisenberg model \cite{wang2015variational}. As previously studied by DMRG
and VMC calculations, $|\text{RVB}\rangle $ describes a
short-range antiferromagnetic ground state, with gapped low-lying
spin excitations\citep{Dagotto1996surprises, Cassanello1996Bilayers}. Based on such $|\text{RVB}\rangle $, we can then optimize
the ground state energies of the RVB state in Eq.~(\ref{rvbgs}) and the non-BCS-like
wavefunction of Eq.~(\ref{gs}) with regard to the variational parameter $%
g(i,j)$. The details of the variational procedure are presented in the Appendix A, which has been developed based on the method firstly applying to the one-hole ground state in Ref. \onlinecite{wang2015variational}.

%%%%%%%%%%%%%%%%%%%%%%%--FIG--%%%%%%%%%%%%%%%%%%%%%%%%%%%%%%%%%%%%%%%%%%
\begin{figure*}[t]
\begin{center}
\includegraphics[width=1\textwidth]{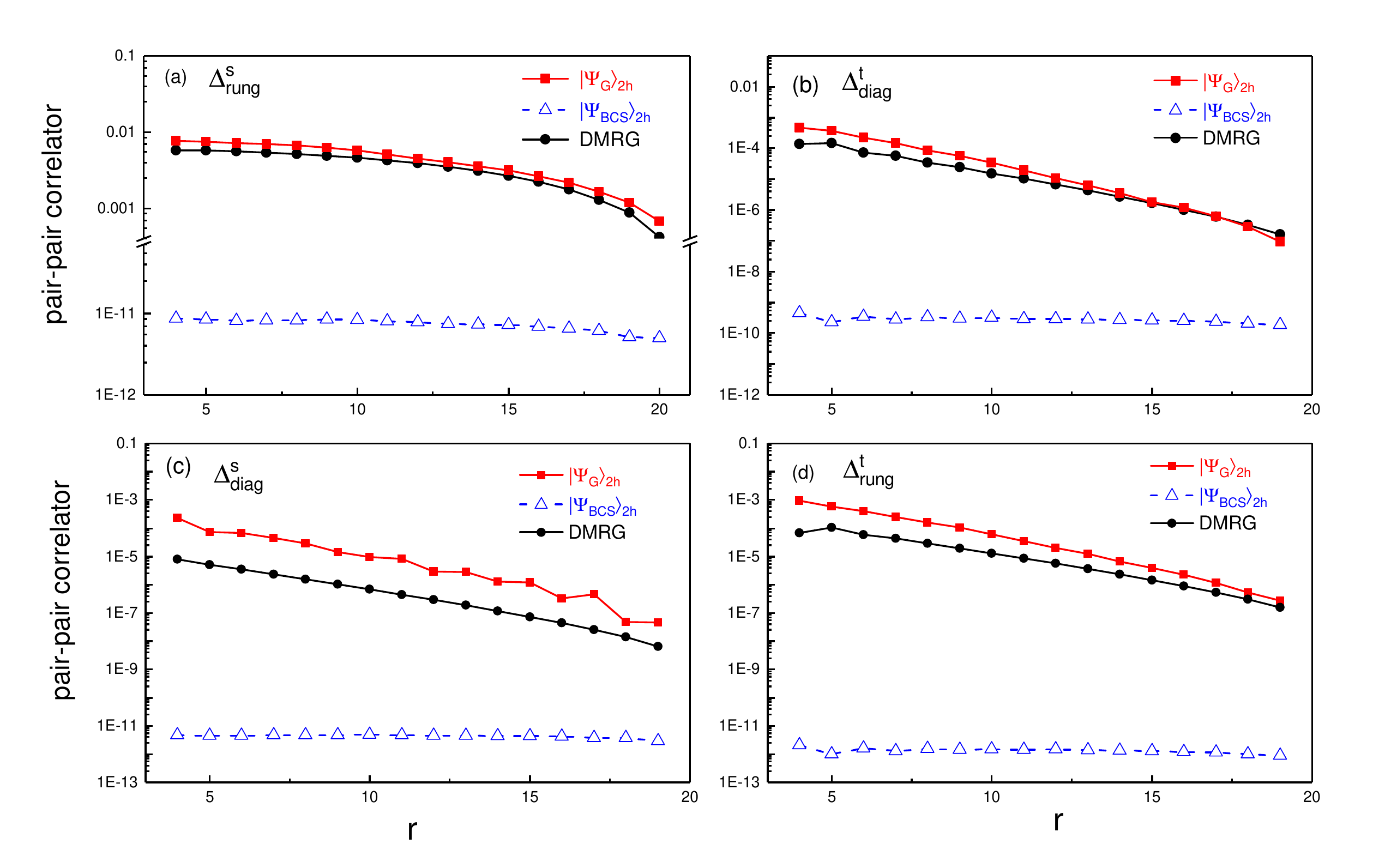}
\end{center}
\par
\renewcommand{\figurename}{Fig.}
\caption{(Color online.) Pair-pair correlators, $C^{s,t}(r)$, as a function of distance $r$  calculated in the ground states $|\Psi _{\mathrm{BCS}}\rangle _{\mathrm{2h}}$ (blue triangle) and $|\Psi _{\mathrm{G}}\rangle _{\mathrm{2h}}$ (red square), respectively, in comparison with the DMRG results (solid circle), are shown for different channels  in (a)-(d) as labeled by the pairing parameters defined in the text. While the present variational ground state  $|\Psi _{\mathrm{G}}\rangle _{\mathrm{2h}}$ in Eq. (\ref{gs}) has well captured the correct pair-pair correlations in all channels as compared with
the DMRG results, the RVB type ground state $|\Psi _{\mathrm{BCS}}\rangle _{\mathrm{2h}}$ in Eq. (\ref{rvbgs}), however, does not show any meaningful pair-pair correlations in all channels. The ladder size is $N=40 \times 2$ with $\protect\alpha=1$.}
\label{pair_cor}
\end{figure*}
%%%%%%%%%%%%%%%%%%%%%%%%%%%%%%%%%%%%%%%%%%%%%%%%%%%%%%%%%%%%%%%%%%%

Figure.~\ref{Energy} shows the variational ground state energies for the two
ground states, Eqs.~(\ref{rvbgs}) and~(\ref{gs}), respectively, as computed
by the VMC method for a finite size ladder. As compared to the DMRG result,
Fig. \ref{Energy} (a) shows that the total energy of the \textquotedblleft
RVB wavefunction\textquotedblright\ in Eq.~\eqref{rvbgs} is indeed much
higher as compared to both the non-BCS wavefunction in Eq.~(\ref{gs}) as well as
the DMRG result, which are relatively much closer. In particular, as shown
in Figs. \ref{Energy} (b) and (c), the deviation between the two variational
ground states mainly comes from the hopping energy $E_{t}=\langle H_t\rangle$, while they agree
well with each other in the superexchange energy $E_{J}=\langle H_J \rangle$. In other words, by
incorporating the non-perturbative phase shift effect in Eqs.~(\ref{tildc}) and (\ref{PS}), the
kinetic energy does get substantially improved in the new ground state (\ref{gs}%
), while the superexchange energy remains approximately unchanged.

Furthermore, an interesting but surprising result is illustrated in Fig. \ref{pair_cor},
in which the pair-pair correlators in the two variational ground states are
computed and compared with the DMRG simulation. For all the spin-singlet and
-triplet channels with the two hole pairing at the same rung or at the
diagonal bond of a plaquette of the ladder in Fig. \ref{Fig1}, the pair-pair correlations are all
vanishingly small ($< 10^{-11}$) in the ``RVB varational wavefunction'' (\ref%
{rvbgs}). By comparison, the pair-pair correlations are much enhanced in the
wavefunction (\ref{gs}) in all channels shown in Fig. \ref{pair_cor}
(a)-(d). In particular the pair-pair correlators of the ground state (\ref{gs}) are in excellent agreement with the DMRG results, which are also shown in Fig. %
\ref{pair_cor} (a)-(d). Note that in Fig. \ref{pair_cor}, the pair-pair
correlators are defined as $C^{s,t}(r)=\left \langle \hat{%
\Delta}_{1i^{\prime }2j^{\prime }}^{s,t}\left( \hat{\Delta}%
_{1i2j}^{s,t}\right) ^{\dagger}\right \rangle $ where the singlet and
triplet channels are 
\begin{equation}
\begin{aligned}  
\hat{\Delta}_{ij}^{s} & =\frac{1}{\sqrt{2}}\sum_{\sigma}\sigma
c^{}_{1i\sigma }c^{}_{2j-\sigma}~, \\
\hat{\Delta}_{ij}^{t} & =\frac{1}{\sqrt{2}}\sum_{\sigma}c^{}_{1i%
\sigma}c^{}_{2j-\sigma}~.
\end{aligned}
\label{Delta}
\end{equation}
Here we only focus on the local rung pairing $\hat{\Delta}_{\text{rung}%
}^{s,t}$ ($i=j$) with $r=|i-i'|$ and local diagonal pairing $\hat{\Delta}_{\text{diag}%
}^{s,t} $($i=j+1$), which represent the dominant pairings in the present two-leg ladder
system\cite{Zhu2017pairing}. 

Therefore, in contrast to the conventional wisdom, the ``RVB ground state" of
Eq.~(\ref{rvbgs}) actually is not in favor of pairing between two holes
upon doping, even though the half-filling $|\text{RVB}\rangle$ state has already
exhibited a short-range antiferromagnetism (an RVB state). On the other hand,
in the new ground state of Eq.~(\ref{gs}), two doped holes do form a strong
bound pair, accompanied by the fact that its kinetic energy is significantly lowered as compared with the variational energy of Eq.~(\ref{rvbgs}). The overall ground state variational energy of $|\Psi_{\mathrm{G}}\rangle_{2\mathrm{h}} $ is in qualitative agreement with the corresponding DMRG result. In particular, the pair-pair correlations calculated based on $|\Psi_{\mathrm{G}}\rangle_{2\mathrm{h}} $ is in excellent agreement with the precise results.
It thus clearly indicates that a \emph{kinetic energy driven mechanism} must be at play in the $t$-$J$ model. This is in sharp contrast to a conventional BCS theory or
Anderson's RVB theory, in which the pairing strength is usually gained from
the potential (superexchange) energy, whereas it causes the further \emph{increase} of the kinetic energy in
forming a bound state. In the following, we further explore the underlying pairing mechanism.

\subsection{Non-BCS pairing mechanism}

The above variational calculations demonstrate that two doped holes injected
into the short-range antiferromagnet $|\text{RVB}\rangle$ can indeed form a tightly bound
state. However, it is not described by Eq.~(\ref{rvbgs}) but by Eq.~(\ref{gs}). The latter is non-BCS-like as each
hole has to simultaneously induce a nonlocal spin ``twist'' via the phase
string factor $e^{-i\hat{\Omega}_i}$ in the spin background $|\text{RVB}%
\rangle$, which is in favor of pairing once two holes are present. By contrast, the pairing between the two holes vanishes in Eq.~(\ref{gs}) once $\hat{\Omega}_i$  is turned off, which results in Eq.~(\ref{rvbgs}).

In order to further understand the underlying physics, let us note that
the two variational states in Eqs.~(\ref{rvbgs}) and (\ref{gs}) can be
connected by the following unitary transformation: 
\begin{equation}  \label{unitary}
|\Psi_{\mathrm{G}}\rangle_{2\text{h}}=e^{i\hat{\Theta}} |
\Psi_{\mathrm{BCS}}\rangle_{2\text{h}}~,
\end{equation}
where 
\begin{equation}\label{duality}
e^{i\hat{\Theta}}\equiv e^{-i\sum\limits_{\gamma i}n_{\gamma
i}^{h}\hat{\Omega}_{\gamma i}}
\end{equation}
with $n_{\gamma i}^{h}$ denoting the hole number operator at the site $i$ of the leg 
$\gamma$ (clearly this unitary transformation can be generalized to arbitrary dopings). 

Then, given the fact that $|\Psi_{\mathrm{G}}\rangle_{2\text{h}}$ is an excellent variational ground state for the $t$-$J$ model,  the ``RVB ground state" $|{{\Psi}_{\mathrm{BCS}}}\rangle_{2\text{h}}$ in Eq.~(\ref{rvbgs}) can be taken as the \emph{correct} trial
wavefunction only if the target Hamiltonian is transformed from the $t$-$J$ type Hamiltonian $H_{t\text{-}J}$ in Eq.~(\ref{t-J}) by ${\widetilde{H}}_{t\text{-}J}\equiv e^{-i\hat{\Theta}}H_{t\text{-}J}e^{i\hat{\Theta}}$, which has the following form 
\begin{equation}  \label{Htld}
{\widetilde{H}}_{t\text{-}J} =H_{\sigma\cdot t\text{-}J}+H_{\mathrm{I}}^{%
\mathrm{string}}~.
\end{equation}
Here the first term on the right-hand-side (rhs) is the $\sigma\cdot t$-$J$
model defined in Sec. II A, in which the hopping term is changed to $H_{\sigma\cdot t}$ in Eq.~(\ref{st}), which is free from the frustration caused by the phase-string sign
structure in the original $t$-$J$ model \citep{Weng2011superconducting, Weng2011mott, Zaanen2011mottness}. It has been previously shown by
DMRG \citep{Zhu2017pairing} that such $\sigma\cdot t$-$J$ model with $%
t_{\perp}=0$ would only lead to a weakly bound state of two holes, which may be
regarded as the RVB mechanism for pairing due to $H_J$. By contrast, the
pairing is absent in the two-leg $\sigma\cdot t$-$J$ ladder model for the isotropic case with $t_{\perp}=t$ \citep{Zhu2014pairing}. However, as pointed out in the above, in either case of $t_{\perp}=0$ or $t_{\perp}=t$, a strong binding
between the two doped holes has been clearly identified in the $t$-$
J$ model by DMRG \citep{Zhu2017pairing,Zhu2014pairing}. 

%%%%%%%%%%%%%%%%%--FIG--%%%%%%%%%%%%%%%%%%%%%%%%%%
\begin{figure}[tbp]
\includegraphics[width=0.5\textwidth]{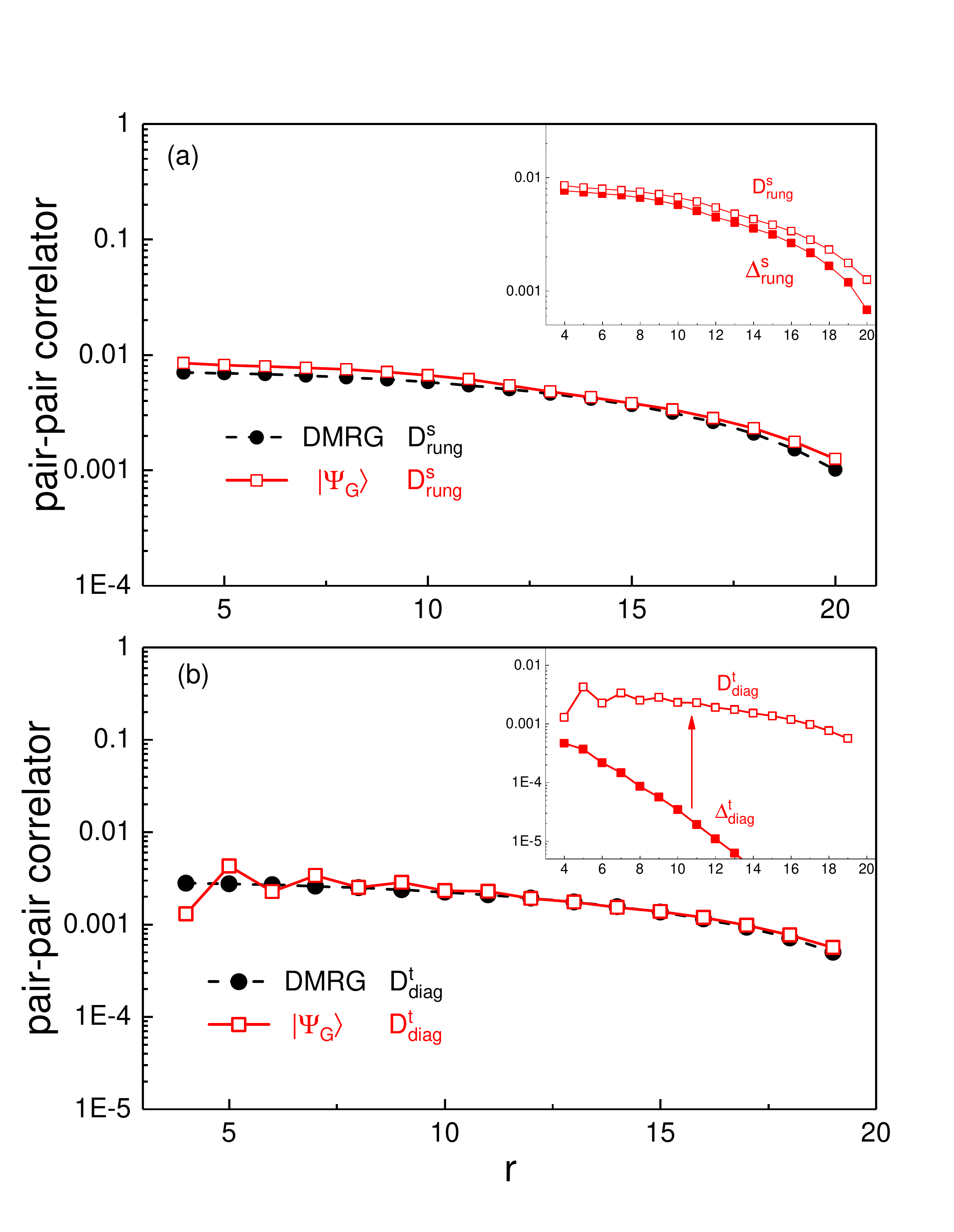}
\par
\renewcommand{\figurename}{Fig.}
\caption{(Color online.) The correlations of $\hat{D}_{\mathrm{diag/rung}}^{s/t}$ defined in Eqs. (\ref{D1}) and (\ref{D2}) calculated by VMC (open square) and DMRG (solid circle) in the singlet (a) and triplet (b) channels. Here  $\hat{D}_{\mathrm{diag/rung}}^{s/t}$ may be regarded as the pair amplitude of the Cooper pair operator $\hat{\Delta}_{\mathrm{diag/rung}}^{s/t}$, and the correlations of both operators are presented in the insets of (a) and (b), which indicate the strengths of the phase fluctuations in the Cooper pair correlators  \citep{Zhu2017pairing}. The system size is $N=40\times2$ with $\protect\alpha=1$.}\label{nonlocal}
\end{figure}
%%%%%%%%%%%%%%%%%%%%%%%%%%%%%%%%%%%%%%%%%%%%%%
Thus, the last term $H_{\mathrm{I}}^{\mathrm{string}}$ on the rhs of Eq.~(\ref{Htld}) in the transformed representation must
play a dominant role in the pairing mechanism of the $t$-$J$ model. It reads \citep{Zhu2017pairing} 
\begin{equation}  \label{HI}
H_{\mathrm{I}}^{\mathrm{string}} =\frac{1}{2} J\sum_{i}(S^{+}_{1i}
S^{-}_{2i}+S^{-}_{1i} S^{+}_{2i})(\Delta\Lambda^{h}_{i}-1),
\end{equation}
where the summation over $i$ is along the chain direction, in which 
\begin{equation}
\Delta\Lambda^{h}_{i}=e^{-i\pi \sum\limits_{l <i}(n^{h}_{1l}-n^{h}_{2l})}
\end{equation}
describes the nonlocal phase shift effect created by the doped holes at both
chains (legs) of $\gamma=1,2$. Since $\langle S^{+}_{1i}
S^{-}_{2i}+S^{-}_{1i} S^{+}_{2i} \rangle <0$ at half-filling, one finds that
two doped holes will generally acquire a string-like pairing potential as follows:

\begin{itemize}
\item If both holes lie on the right hand or left hand of the rung $1i,2i$,
the factor $\Delta \Lambda^{h}_{i}=1$ makes a vanishing contribution in Eq.~(%
\ref{HI}).

\item Only when the rung $1i,2i$ is sandwiched by the two holes along the
chain direction, does the factor $\Delta \Lambda^{h}_{i}=-1$ make a finite
contribution in Eq. \eqref{HI}.
\end{itemize}
Consequently, an effective potential given by Eq.~(\ref{HI}) for two holes
can be found 
\begin{equation}\label{V}
V(h_1,h_2)\varpropto J|{x}_{h_1}-x_{h_2}|~,
\end{equation}
where  $|{x}_{h_1}-x_{h_2}|$ denotes the distance between the two holes at site 
$h_1$ and $h_2$ along the chain ($x$) direction.

Namely, if one insists on using the BCS-type wavefunction of Eq.~(\ref{rvbgs}) to describe the hole pairing ground state, then the original $t$-$J$ Hamiltonian has to be transformed into a new Hamiltonian ${\widetilde{H}}_{t\text{-}J}$ in Eq.~(\ref{Htld}), in which the hopping term is replaced by that of the $\sigma\cdot t$-$J$ that is free from the phase string effect. Nevertheless, there emerges an additional nonlocal string-like pairing potential besides the original superexchange term. It is this new string-like potential $H_{\mathrm{I}}^{\mathrm{string}}$ that will lead to the strong binding between the two doped holes in ${\widetilde{H}}_{t\text{-}J}$  rather than the superexchange term $H_J$ in the $\sigma\cdot t$-$J$ term in Eq.~(\ref{Htld}).   

Let us further examine the pair-pair correlators in such a transformed
representation. Note that in the new Hamiltonian (\ref{Htld}) the pair
operators $\hat{\Delta}_{ij}^{s,t}$ defined in Eq.~(\ref{Delta}) will correspond to the following operators $%
\hat{D}_{ij}^{s,t}$ in the original $t$-$J$ model: 
\begin{align}\label{D1}
\hat{D}_{ij}^{s,t}&\equiv e^{i\hat{\Theta}}\hat{\Delta}_{ij}^{s,t} 
e^{-i\hat{\Theta}} \nonumber\\
&=\hat{\Delta}_{ij}^{s,t}e^{-i\left(\hat{\Omega}_i+\hat{\Omega}_j\right)}~,
\end{align}
or 
\begin{equation}\label{D2}
\begin{aligned}  
\hat{D}_{ij}^{s} & =\frac{1}{\sqrt{2}}\sum_{\sigma}\sigma \tilde{c}%
^{}_{1i\sigma }\tilde{c}^{}_{2j-\sigma}~, \\
\hat{D}_{ij}^{t} & =\frac{1}{\sqrt{2}}\sum_{\sigma}\tilde{c}^{}_{1i\sigma}%
\tilde{c}^{}_{2j-\sigma}~.
\end{aligned}
\end{equation}
One may then calculate the pair-pair correlators of $\hat{D}_{ij}^{s,t}$ based on Eq.~(\ref{D1}) in the original representation of $H_{t\text{-}J} $ and $|\Psi_{\mathrm{G}%
}\rangle_{2\text{h}}$.
As shown in Fig.~\ref{nonlocal}, the VMC calculations are in excellent agreement with the DMRG
simulation, indicating that the two-hole ground state in the transformed representation governed by the new Hamiltonian (\ref{Htld}) is indeed described by a BCS-like ``Cooper pairing" in the wavefunction (\ref{rvbgs}). Equivalently in the original representation, it is the operator $\tilde{c}$ instead of the bare hole creation operator $c$ that plays the central role in ``Cooper pairing".  

As indicated in the insets of Figs.~\ref{nonlocal} (a) and (b), the strengths of the $\hat{D}$ correlators
generally get enhanced as compared with those of the true Cooper pairs characterized by $\hat{\Delta}$ in the original $t$-$J$ model. It indicates that the \emph{true} Cooper pair operator $\hat{\Delta}$
must possess a composite structure including both a pairing amplitude (mean-field-like) $\hat{D}
_{ij}^{s,t}$ and a phase fluctuation as shown in Eq.~(\ref{D1}), which has already been established by the DMRG calculation in Ref. \onlinecite{Zhu2017pairing}.

\section{Discussion}

In this work, the pairing of two holes doped into a Mott insulator has been studied by the VMC calculation. Specifically, we have explored a non-BCS-type wavefunction [Eq.~(\ref{gs})] with incorporating an intrinsic phase fluctuation discovered in a previous DMRG approach \citep{Zhu2014pairing,Zhu2017pairing}. Such a new variational wavefunction has been shown to give rise to the correct behavior of the two-hole ground state in comparison with the DMRG results \citep{Zhu2017pairing}. By contrast, the conventional BCS (RVB) like wavefunction [Eq.~(\ref{rvbgs})] leads to the qualitatively wrong behavior in both the ground state energy and the pair-pair correlations. In particular it predicts the absence of any meaningful pairing as opposed to a strong binding between the holes as revealed by DMRG \citep{Zhu2017pairing} as well as by the present variational wavefunction. 

It means that the so-called RVB mechanism \cite{Anderson1987} is not sufficient at least in the present two-leg ladder case in describing the hole pairing, even though the spin-spin correlation is already short-ranged here as envisaged originally by Anderson \cite{Anderson1987} for an RVB state. Rather than the conventional RVB pairing potential contributed by the superexchange term $H_J$, the strong pairing state for the two holes is found to be due to a distinct mechanism, which is ``kinetic energy driven'' by nature. Namely, each doped hole will have to induce an irreparable phase string effect in the spin background \citep{Weng2011superconducting, Weng2011mott, Zaanen2011mottness}, which strongly frustrates its kinetic energy  \citep{Zhu2014pairing,Zhu2017pairing}.  The bare hole is then renormalized by a nonlocal phase shift to form a ``twisted'' quasiparticle as given in Eqs.~(\ref{tildc}) and (\ref{PS}). What we have found in this work is that two of the twisted holes can indeed form a tightly bound pair as described by Eq.~(\ref{gs}), and by doing so the strong frustration on the kinetic energy can be effectively released.
  
Thus, the Cooper pairing of two doped holes can no longer be simply attributed to exchanging a ``bosonic mode'' or via an RVB pairing of the spins. Instead, the dominant pairing force is originated from the phase string effect of the $t$-$J$ model. Such a non-BCS pairing force can be explicitly revealed by utilizing a unitary transformation to  ``gauge away'' the phase string effect from the hopping term, which results in an effective string-like pairing potential in Eq.~(\ref{HI}) that is nonlocal and of non-perturbative nature. Physically, the phase string effect can be also interpreted in terms of the spin current backflow produced by the hopping of the doped holes\citep{Zheng2018hidden}. In this sense, the string-like force shown in Eqs.~(\ref{HI}) and (\ref{V}) may be also regarded as a special type of the Amperean pairing potential \cite{Lee2014amperean,Lee2007amperean}.  

We point out that both the ansatz states given in Eqs.~(\ref{gs}) and (\ref{rvbgs}) have omitted the usual ``spin-polaron''\citep{Schmitt1988spectral, Kane1989motion,Martinez1991spin, Liu1991spectral} or ``spin bag'' effect\citep{Schrieffer1988spin, Weng1988d}, which arises from the ``amplitude'' distortion in the spin background around the doped hole, in contrast to the ``phase'' or the ``transverse'' (spin current) distortion given in Eq.~(\ref{tildc}). The former should further improve the variational ground state energy shown in Fig.~\ref{Energy}, and renormalize the effective mass of the doped hole. But we do not think such an effect will violate the Landau's one-to-one correspondence, as the present phase shift in Eq.~(\ref{tildc}) does, to result in a qualitative change in the ground state properties, including the pairing mechanism. Nevertheless, properly including such an effect is expected to further lower the variational energy of Eq.~(\ref{gs}) in comparison with the DMGR results, even though the pair-pair correlations should not be improved much according to Fig. \ref{pair_cor}.

The present study of the hole pairing in the $t$-$J$ model has been carried out in one of the simplest limits. Namely, we have considered two holes
doped into a spin gapped two-leg Heisenberg ladder, in which two holes are only allowed to hop along the chain direction of the ladder with $t_{\perp}=0$. As the consequence, the phase shift operator $\hat{\Omega}_i$ takes the simple one-dimensional form given in Eq.~(\ref{PS}). On the other hand, with $t_{\perp}\neq 0$, the DMRG calculation (cf. Appendix B) shows that the two-hole ground state persists continuously from $t_{\perp}= 0$ without phase transition. In other words, the non-BCS pairing revealed in the present work should remain similar at $t_{\perp}\neq 0$. There, the form of $\hat{\Omega}_i$ associated with one hole doping will generally involve both two chains of the two-leg ladder as previously shown in Ref.~\onlinecite{wang2015variational}. The pairing of the two twisted holes should thus remain the same as in Eq.~(\ref{gs}) in the variational approach, with $\hat{\Omega}_i$ being modified. A similar approach may be further generalized to the two-dimensional case, where the phase shift operator $\hat{\Omega}_i$ will take an isotropic form\citep{Weng2011superconducting, Weng2011mott, Zaanen2011mottness}. 

Finally, a natural generalization of the ground state ansatz in Eq.~(\ref{gs}) for the two-hole case to the finite doping may be straightforwardly written down as follows, 
\begin{equation}  \label{ggs}
|\Psi_{\mathrm{G}}\rangle=e^{\hat {D}} |\text{RVB}\rangle~,
\end{equation}
which has been previously constructed in Ref.~\onlinecite{Weng2011superconducting}, where $|\text{RVB}\rangle$ still denotes a spin ``vacuum'' state and the
``twisted'' Cooper pair $\hat {D}$ is defined in Eq.~(\ref{D}). As a technical remark, we note here that the compact form in Eq.~(\ref{ggs}) should be correctly understood as an abbreviation expression for a truly fractionalized state \citep{Weng2011mott, Weng2011superconducting}. That is, the phase shift operator $\hat{\Omega}_i$ in $\hat {D}$ [Eq.(\ref{D})] should always act on the half-filling vacuum state $|\text{RVB}\rangle$ \emph{before} the annihilations of the electrons at the hole sites by $\hat {D}$, which can only be precisely implemented by introducing a specific fractionalization  \citep{Weng2011mott, Weng2011superconducting}. By such a construction, the pairing amplitude $g(i,j)$ in Eq.~(\ref{PS}) and the RVB
pairing in $|\text{RVB}\rangle$ can still remain mean-field-like to give rise to a nontrivial/non-BCS superconducting ground state at finite doping, which is to be further investigated variationally elsewhere.

\section{Acknowledgements}

Useful discussions with Qing-Rui Wang, Yang Qi, D.N. Sheng are acknowledged. S. C. is indebted to Si-Bo Zhao for the help with computation. This work is partially supported by Natural Science Foundation of China (Grant No. 11534007), and MOST of China (Grants No. 2015CB921000 and No. 2017YFA0302902).

\appendix
\newpage \onecolumngrid

\section{Variational Monte Carlo procedure}

For the sake of self-consistency, we first present the VMC procedures for
the half-filled RVB state. Subsequently, we derive the two-hole variational
wavefunctions and some formulas used in the VMC procedure.

\subsection{VMC for half-filling wavefunctions}

At half-filling both the $t$-$J$ and $\sigma \cdot t$-$J$ model reduce to a
pure Heisenberg spin ladder whose ground state $|$RVB$\rangle $ is an
anti-ferromagnetic gapped system. A pure Heisenberg model can have a good
simulation by the Liang-Doucot-Anderson type bosonic RVB variational
wavefunction\citep{Liang1988some} :%
\begin{equation}
|\text{RVB}\rangle =\sum_{\upsilon }\omega _{\upsilon }|\upsilon \rangle ~,
\label{RVB}
\end{equation}%
where $|\upsilon \rangle $ is a singlet pairing valence bond (VB) state
where spins on sites $i$ and $j$ from different sublattices form a singlet
pairing, which enables $|\mathrm{RVB}\rangle $ to obey the Marshall sign rule\citep{marshall1955antiferromagnetism}.
The amplitude of each VB state $|\upsilon \rangle $ can be factorized by $%
\omega _{\upsilon }=\prod_{\left( ij\right) \in \upsilon }h_{ij}$. Here $%
h_{ij}$ is a non-negative function with respect to sites $i$ and $j$. Such a
scheme will tremendously decrease the number of variational parameters.
In Ref.~\onlinecite{wang2015variational}, the VMC calculations for a $40\times2$ Heisenberg ladder system with open boundary condition show high accuracy as compared with the DMRG results.

The norm of the RVB state in Eq.~(\ref{RVB}) is given as 
\begin{equation}
\left\langle \text{RVB}|\text{RVB}\right\rangle =\sum_{\upsilon ,\upsilon
^{\prime }}\omega _{\upsilon }\omega _{\upsilon ^{\prime }}\left\langle
\upsilon ^{\prime }|\upsilon \right\rangle ~.  \label{norm}
\end{equation}%
The positiveness of $\omega _{\upsilon }\omega _{\upsilon ^{\prime
}}\left\langle \upsilon ^{\prime }|\upsilon \right\rangle $ allows an
explanation as a distribution function. The sampling of $\left\langle
\upsilon ^{\prime }|\upsilon \right\rangle $ is time-consuming. We can
introduce the Ising configuration $\sigma $ (simply use $\sigma $ for $%
\left\{ \sigma \right\} $), whose relation to the VB state is 
\begin{equation}
|\sigma \rangle \left\langle \sigma |\upsilon \right\rangle =\delta
_{\upsilon ,\sigma }|\upsilon ,\sigma \rangle ~,
\end{equation}%
in which $\delta _{\upsilon ,\sigma }=\left\vert \left\langle \sigma
|\upsilon \right\rangle \right\vert $ and $\left\langle \sigma |\upsilon
\right\rangle $ is zero or the Marshall sign for the RVB state. Then the RVB
state in Eq.~(\ref{RVB}) can be rewritten as 
\begin{equation}
|\text{RVB}\rangle =\sum_{\upsilon }\omega _{\upsilon }|\upsilon \rangle
=\sum_{\upsilon ,\sigma }\delta _{\upsilon ,\sigma }|\upsilon ,\sigma
\rangle ~.
\end{equation}%
The summation is constrained in the space where the VB state $|\upsilon
\rangle $ is compatible with the Ising basis $|\sigma \rangle $. With the
fact 
\begin{align*}
\left\langle \upsilon ^{\prime }|\upsilon \right\rangle & =2^{N_{\upsilon
,\upsilon ^{\prime }}^{\text{loop}}}, \\
\left\langle \upsilon ^{\prime },\sigma ^{\prime }|\upsilon ,\sigma
\right\rangle & =\delta _{\sigma ,\sigma ^{\prime }},
\end{align*}%
the norm in Eq.~(\ref{norm}) has a more explicit form%
\begin{equation}
\left\langle \text{RVB}|\text{RVB}\right\rangle =\sum_{\upsilon ,\upsilon
^{\prime },\sigma }\omega _{\upsilon }\omega _{\upsilon ^{\prime }}\delta
_{\upsilon ^{\prime },\sigma }\delta _{\upsilon ,\sigma }~.
\end{equation}%
Here $N_{\upsilon ,\upsilon ^{\prime }}^{\text{loop}}$ is the number of
loops in the transposition-graph covers $\left( \upsilon ,\upsilon ^{\prime
}\right) $.

The formulas for averaging physical operators can be found in Ref.~%
\onlinecite{wang2015variational}. Whereafter, we will generalize the same
trick to two-hole wavefunctions. 

\subsection{Two-hole ground state}

We can construct a two-hole VB state by removing two electrons with opposite
spin indexes from the half-filled VB state:%
\begin{equation}
|h_{1},h_{2},\upsilon \rangle \equiv \mathrm{sgn}(h_{1}-h_{2})c_{h_{1}%
\uparrow }^{{}}c_{h_{2}\downarrow }^{{}}|\upsilon \rangle =\sum_{\sigma
_{h}}\delta _{\upsilon ,\sigma _{h}}|h_{1},h_{2},\sigma _{h}\rangle ~,
\end{equation}%
where $|\upsilon \rangle $ is a half-filled VB state and $%
|h_{1},h_{2},\sigma _{h}\rangle \equiv \text{sgn}(h_{1}-h_{2})c_{h_{1}^{{}}%
\uparrow }^{{}}c_{h_{2}^{{}}\downarrow }^{{}}|\sigma _{h}\rangle $ with $%
|\sigma _{h}\rangle $ denoting a half-filled Ising basis. The function $%
\mathrm{sgn}(h_{1}-h_{2})$ is the sign function i.e if $h_{1}>h_{2},$ $\mathrm{sgn}%
(h_{1}-h_{2})=1$; if $h_1=h_2$, sign$(h_1-h_2)=0$  and if $h_{1}<h_{2},$ $\mathrm{sgn}(h_{1}-h_{2})=-1$. 
 If $\upsilon $ and $\sigma _{h}$ are not
compatible, $\delta _{\upsilon ,\sigma _{h}}=0$, namely for some dimmer $\left( i,j\right) ,$ $\sigma _{h}\left(
i\right) =\sigma _{h}\left( j\right) $, ($\sigma _{h}\left( i\right) $ is
the spin index on the site $i$ in the Ising basis $|\sigma _{h}\rangle $) or 
$\sigma _{h}\left( h_{1}\right) =\uparrow $ or $\sigma _{h}\left(
h_{2}\right) =\downarrow $.

The two-hole variational wavefunction is obtained by removing two
anti-directed spins on the RVB state in Eq.~(\ref{RVB}) accompanied with a
unitary transformation $\hat{\Lambda}$%
\begin{align}
|\Psi \rangle _{\mathrm{G}}& =\hat{\Lambda}\sum_{\substack{ h_{1},h_{2} \\ %
h_{1}\neq h_{2}}}g(h_{1},h_{2})\text{sign}(h_{1}-h_{2})c_{h_{1}\uparrow
}^{{}}c_{h_{2}\downarrow }^{{}}|\text{RVB}\rangle   \notag \\
& =\sum_{\substack{ h_{1},h_{2} \\ h_{1}\neq h_{2}}}\sum_{\sigma
_{h},\upsilon }g(h_{1},h_{2})\hat{\Lambda}\left( h,\sigma _{h}\right) \delta
_{\upsilon ,\sigma _{h}}\omega _{\upsilon }|h_{1},h_{2},\sigma _{h}\rangle 
\label{wf}
\end{align}%
in which $g(h_{1},h_{2})$ is the hole wavefunction that only depend on
holes' position and it will entangle with antiferromagnetic background
through the phase operator $\hat{\Lambda}$. 
The phase $\Lambda (h,\sigma _{h})$ generally is the function of the two
hole positions $h_{1}$ and $h_{2}$ and spin configuration $\sigma _{h}$ and
is defined by 
\begin{equation}
\hat{\Lambda}\left( h,\sigma _{h}\right) |h_{1},h_{2},\sigma _{h}\rangle
=\prod_{h\in \left\{ h_{1},h_{2}\right\} }\prod_{l\neq h_{1},h_{2}}\Lambda
\left( h,l,\sigma _{h}\left( l\right) \right) |h_{1},h_{2},\sigma
_{h}\rangle   \label{phasefactorOfaSite}
\end{equation}%
We factorize $\hat{\Lambda}\left( h,\sigma _{h}\right) $ via $\Lambda \left(
h,l,\sigma _{h}\left( l\right) \right) $, which is a phase factor felt by a
hole from the spin at the site $l$. Specifically, it has different forms for
different variational assumptions:

\begin{enumerate}[ 1)]
\item If we take $\Lambda \left( h,l,\sigma _{h}\left( l\right) \right) =1$,
then $\hat{\Lambda}=1$ and we get the BCS-type wavefunction in Eq.~(1),

\item For the non-BCS type wavefunction in Eq.~(4) in the $t$-$J$
ladder system with $t_{\perp }=0$ in the main body, 
\begin{equation}
\Lambda \left( h,l,\sigma _{h}\left( l\right) \right) =\left\{ 
\begin{array}{cl}
            1                                                                                    &  h,l\mbox{ lie in different legs}\\
 e^{-i\pi }\delta _{\sigma _{h}\left( l\right) ,\downarrow } &h,l\mbox{ lie in the same leg and }x_{l}>x_{h} \\ 
          1                        &h,l\mbox{ lie in the same leg and }  x_{l}<x_{h}%
\end{array}%
\right. 
\end{equation}%
where $x_l$ is the coordinate of site $l$ along the chain ($x$) direction.
\item For $t$-$J$ model with $t_{\perp }\not=0$ ladder systems or 2D
systems, the expression of $\Lambda \left( h,l,\sigma _{h}\left( l\right)
\right) $ can be found in Ref.~\onlinecite{wang2015variational}.
\end{enumerate}

With the inner product formulas%
\begin{align}
\langle h_{1}^{\prime },h_{2}^{\prime },\sigma _{h^{\prime }}^{\prime
}|h_{1},h_{2},\sigma _{h}\rangle & =\delta _{h_{1}^{\prime },h_{1}}\delta
_{h_{2}^{\prime },h_{2}}\delta _{\sigma _{h},\sigma _{h}^{\prime }}\delta
_{\sigma _{h}\left( h_{1}\right) ,-\sigma _{h}\left( h_{2}\right) }, \\
\langle h_{1}^{\prime },h_{2}^{\prime },\upsilon ^{\prime
}|h_{1},h_{2},\upsilon \rangle & =\left\{ 
\begin{array}{ll}
\delta _{h_{1}^{\prime },h_{1}}\delta _{h_{2}^{\prime },h_{2}}2^{N_{\upsilon
,\upsilon ^{\prime }}^{\text{loop}}-1}\left( 1-\delta _{h_{1}h_{2}}^{\text{%
sublatt}}\right) & h_{1},h_{2}\in \text{s.l. } \\ 
\delta _{h_{1}^{\prime },h_{1}}\delta _{h_{2}^{\prime },h_{2}}2^{N_{\upsilon
,\upsilon ^{\prime }}^{\text{loop}}-2} &   h_{1},h_{2}\notin \text{s.l.}%
\end{array}%
\right. 
\end{align}%
where $h_{1},h_{2}\in $s.l. means that sites $h_{1}$ and $h_{2}$ belong to
the same close loop in the transposition graph $\left( \upsilon ,\upsilon
^{\prime }\right) $ and $\delta _{h_{1}^{\prime },h_{1}}=1$ if $%
h_{1}^{\prime }=h_{1}$, and otherwise $\delta _{h_{1}^{\prime },h_{1}}=0.$ $%
\delta _{h_{1}h_{2}}^{\text{sublatt}}=1$ if sites $h_{1},h_{2}$ are in the
same sublattice, and otherwise, $\delta _{h_{1}h_{2}}^{\text{sublatt}}=0,$
we can express the norm of $|\Psi \rangle _{\mathrm{G}}$ as :%
\begin{equation}
\langle \Psi |\Psi \rangle _{\mathrm{G}}=\frac{1}{4}\sum_{\upsilon ,\upsilon
^{\prime },\sigma ^{0}}\delta _{\upsilon ,\sigma ^{0}}\delta _{\upsilon
^{\prime },\sigma ^{0}}\omega _{\upsilon ^{\prime }}\omega _{\upsilon }\left[
\sum_{\substack{ h_{1},h_{2}, \\ h_{1},h_{2}\in \text{s.l.}}}2\left(
1-\delta _{h_{1}h_{2}}^{\text{sublatt}}\right) \text{\ }+\sum_{\substack{ %
h_{1},h_{2}, \\ h_{1},h_{2}\notin \text{s.l.}}}1\right] \left\vert
g(h_{1},h_{2})\right\vert ^{2},
\end{equation}%
where $\sigma ^{0}$ is a compatible spin configuration with a transposition
graph $\left( \upsilon ,\upsilon ^{\prime }\right) .$ Note that the norm of $%
|\Psi \rangle _{\mathrm{G}}$\ depends on different VB configuration $\left(
\upsilon ,\upsilon ^{\prime }\right) .$ To overcome it, we have to employ a
similar trick as Gutzwiller projection\citep{edegger2007gutzwiller}: using average values of $\left[ \sum
_{\substack{ h_{1},h_{2}, \\ h_{1},h_{2}\in \text{s.l.}}}2\left( 1-\delta
_{h_{1}h_{2}}^{\text{sublatt}}\right) \ +\sum_{\substack{ h_{1},h_{2}, \\ %
h_{1},h_{2}\notin \text{s.l.}}}1\right] \left\vert g(h_{1},h_{2})\right\vert
^{2}$ under the half-filled RVB state instead: 
\begin{align}
& \left\langle \left[ \sum_{\substack{ h_{1},h_{2}, \\ h_{1},h_{2}\in \text{%
s.l.}}}2\left( 1-\delta _{h_{1}h_{2}}^{\text{sublatt}}\right) \ +\sum
_{\substack{ h_{1},h_{2}, \\ h_{1},h_{2}\not\in \text{s.l.}}}1\right]
\left\vert g(h_{1},h_{2})\right\vert ^{2}\right\rangle _{\text{RVB}}  \notag
\\
& =\sum_{h_{1},h_{2}}\left[ 2\left( 1-\delta _{h_{1}h_{2}}^{\text{sublatt}%
}\right) P_{h_{1}h_{2}}+\left( 1-P_{h_{1}h_{2}}\right) \right] \left\vert
g(h_{1},h_{2})\right\vert ^{2}  \notag \\
& \equiv \sum_{h_{1},h_{2}}a\left( h_{1},h_{2}\right) ^{2}\left\vert
g(h_{1},h_{2})\right\vert ^{2},
\end{align}%
where $P_{h_{1}h_{2}}$ describes the possibility of two sites $h_{1},h_{2}$
belonging to the same loop in all the transposition graphs $\left( \upsilon
,\upsilon ^{\prime }\right) $. The factor $a(h_{1},h_{2})^{2}\equiv 2\left(
1-\delta _{h_{1}h_{2}}^{\text{sublatt}}\right) P_{h_{1}h_{2}}+\left(
1-P_{h_{1}h_{2}}\right) $ will regularize the norm that relates $|\Psi
\rangle _{\text{G}}$ to $|\text{RVB}\rangle $: 
\begin{equation}
\sum_{\substack{ h_{1},h_{2} \\ h_{1}\not=h_{2}}}a\left( h_{1},h_{2}\right)
^{2}\left\vert g(h_{1},h_{2})\right\vert ^{2}=1.  \label{renorm-g}
\end{equation}%
\begin{equation}
\langle \Psi |\Psi \rangle _{\mathrm{G}}=\frac{1}{4}\sum_{\upsilon ,\upsilon
^{\prime },\sigma ^{0}}\delta _{\upsilon ,\sigma ^{0}}\delta _{\upsilon
^{\prime },\sigma ^{0}}\omega _{\upsilon ^{\prime }}\omega _{\upsilon }=%
\frac{1}{4}\left\langle \text{RVB}|\text{RVB}\right\rangle 
\label{reduced norm}
\end{equation}%
In Sec~\ref{Sup::VMCPro}, we will describe the procedures for variational
optimization on the wavefunction $g(h_{1},h_{2})$. Together with Eq.~%
\eqref{wf} and the identity $\langle \upsilon |\upsilon ^{\prime }\rangle
=\sum_{\sigma ^{0}}\delta _{\upsilon ,\sigma ^{0}}\delta _{\upsilon ^{\prime
},\sigma ^{0}}$, the expectation value of an operator $\hat{O}$ can be
generally expressed as 
\begin{align}
\langle \hat{O}\rangle & =\frac{\langle \Psi |\hat{O}|\Psi \rangle _{\mathrm{%
G}}}{\langle \Psi |\psi \rangle _{\mathrm{G}}}  \notag \\
& =\frac{4\left( \sum_{\upsilon ,\upsilon ^{\prime },\sigma ^{0}}\delta
_{\upsilon ,\sigma ^{0}}\delta _{\upsilon ^{\prime },\sigma ^{0}}\right)
\omega _{\upsilon ^{\prime }}\omega _{\upsilon }\sum_{h_{1}^{\prime
}\not=h_{2}^{\prime },\sigma _{h^{\prime }}^{\prime },\sigma _{h}}\delta
_{\upsilon ^{\prime },\sigma _{h^{\prime }}^{\prime }}\delta _{\upsilon
,\sigma _{h}}E\left( \hat{O}\right) }{\left( \sum_{\upsilon ,\upsilon
^{\prime },\sigma ^{0}}\delta _{\upsilon ,\sigma ^{0}}\delta _{\upsilon
^{\prime },\sigma ^{0}}\right) \omega _{\upsilon ^{\prime }}\omega
_{\upsilon }}  \label{MC mother}
\end{align}%
where 
\begin{equation}
E(\hat{O})=\mathrm{Re}(\Delta \Lambda )\frac{\langle h_{1}^{\prime
},h_{2}^{\prime },\sigma _{h^{\prime }}^{\prime }|O|h_{1},h_{2},\sigma
_{h}\rangle }{\langle \upsilon ^{\prime }|\upsilon \rangle }
\end{equation}%
and 
\begin{equation}
\mathrm{Re}(\Delta \Lambda )=\mathrm{Re}\left[ \Lambda ^{\ast
}(h_{1}^{\prime },h_{2}^{\prime },\sigma _{h^{\prime }}^{\prime })\Lambda
(h_{1},h_{2},\sigma _{h})g^{\ast }(h_{1}^{\prime },h_{2}^{\prime
})g(h_{1},h_{2})\right] .
\end{equation}%
Here $\mathrm{Re}(\Delta \Lambda )$ denotes the real part of $\Delta \Lambda 
$.\ We interpret $\omega _{\upsilon ^{\prime }}\omega _{\upsilon }/\langle
\Psi |\Psi \rangle _{\mathrm{G}}$ as a distribution function in the space of
compatible spin configurations $\left( \upsilon ,\upsilon ^{\prime },\sigma
^{0}\right) .$ 

\subsection{VMC procedure}

\label{Sup::VMCPro}

We have to optimize parameters $h_{ij}$ of the background RVB\citep{sandvik2010loop} and the
wavefunction $g\left( h_{1},h_{2}\right) $ with respect to the total energy $%
E_{\mathrm{total}}$. 
%\textbf{(As mentioned in the main text, for the gapped pure Heisenberg
%spin ladder with a physical }$t/J$\textbf{\ value, the spin polaron will
%only renormalize the effective mass of quasiparticles. Thus, its effect can
%be neglected in the variational wavefunctions (\ref{wf}).)}
The total energy of the system reads 
\begin{equation}
E_{\mathrm{total}}=\left\langle \Psi \left\vert H_{t}+H_{J}\right\vert \Psi
\right\rangle _{\text{G}}=%
\sum_{j_{1}j_{2},i_{1}i_{2}}H_{j_{1}j_{2},i_{1}i_{2}}^{\text{eff}}g^{\ast
}(j_{1},j_{2})g(i_{1},i_{2})~,
\end{equation}%
where $H^{\text{eff}}$ is an effective Hamiltonian for the hole wavefunction 
$g$, 
\begin{equation}
H^{\text{eff}}=H_{t}^{\text{eff}}+H_{J}^{\text{eff}}~.
\end{equation}%
We introduce a renormalized wavefunction $\tilde{g}$ to incorporate with its
normalization condition (\ref{reduced norm}),%
\begin{equation*}
\tilde{g}\left( h_{1},h_{2}\right) =a\left( h_{1},h_{2}\right) g\left(
h_{1},h_{2}\right) ~.
\end{equation*}%
Consistently, $H^{\text{eff}}$ will be transformed into $\widetilde{H}^{%
\text{eff}}$ whose elements are 
\begin{equation}
\widetilde{H}_{j_{1}j_{2},i_{1}i_{2}}^{\text{eff}}=a^{-1}\left(
j_{1},j_{2}\right) H_{j_{1}j_{2},i_{1}i_{2}}^{\text{eff}}a^{-1}\left(
i_{i},i_{2}\right) ~.
\end{equation}%
Thus the total energy is expressed as 
\begin{equation}
E_{\mathrm{total}}=\sum_{j_{1}j_{2},i_{1}i_{2}}\widetilde{H}%
_{j_{1}j_{2},i_{1}i_{2}}^{\text{eff}}\tilde{g}^{\ast }(j_{1},j_{2})\tilde{g}%
(i_{1},i_{2})~,
\end{equation}%
with the normalization condition for $\tilde{g}(i_{1},i_{2})$: 
\begin{equation}
\sum_{i_{1},i_{2}}|\tilde{g}(i_{1},i_{2})|^{2}=1~.
\end{equation}%
Diagonalize $\widetilde{H}^{\text{eff}}$ and the minimal eigenvalue and the
corresponding eigenvector are the variational energy and renormalized
wavefunction respectively. All remaining are to simulate elements of $H_{t}^{%
\text{eff}}$ and $H_{J}^{\text{eff}}$. In the following, we provide some
explicit formulas used in the Monte Carlo simulation.

\subsection{Superexchange Energy}

Superexchange terms $H_{J}$ only affect spin configuration with the
positions of holes unchanged, which will simplify calculation processes. The
average value of the superexchange energy of two sites $i,j$ with fixed
positions of holes is 
\begin{equation}
\langle H_{ij}^{J}\rangle =\frac{(\sum_{\upsilon ,\upsilon ^{\prime },\sigma
^{0}}\delta _{\upsilon ,\sigma ^{0}}\delta _{\upsilon ^{\prime },\sigma
^{0}})\omega _{\upsilon ^{\prime }}\omega _{\upsilon }E_{ij}^{J}\left(
\upsilon ,\upsilon ^{\prime }\right) }{(\sum_{\upsilon ,\upsilon ^{\prime
},\sigma ^{0}}\delta _{\upsilon ,\sigma ^{0}}\delta _{\upsilon ^{\prime
},\sigma ^{0}})\omega _{\upsilon ^{\prime }}\omega _{\upsilon }}~,
\end{equation}%
where 
\begin{align}
E_{ij}^{J}\left( \upsilon ,\upsilon ^{\prime }\right) & =\left\vert
g(h_{1},h_{2})\right\vert ^{2}E_{ij}^{J}\left( h,\upsilon ,\upsilon ^{\prime
}\right) , \\
E_{ij}^{J}\left( h,\upsilon ,\upsilon ^{\prime }\right) & =\sum_{\sigma
_{h^{\prime }}^{\prime }\sigma _{h}^{{}}}\delta _{\upsilon ^{\prime },\sigma
_{h^{\prime }}^{\prime }}\delta _{\upsilon ,\sigma _{h}^{{}}}\mathrm{Re}%
\left( \Lambda ^{\ast }(h_{1}^{{}},h_{2}^{{}},\sigma _{h}^{\prime })\Lambda
(h_{1}^{{}},h_{2}^{{}},\sigma _{h})\right) \frac{4\langle h_{1},h_{2},\sigma
_{h}^{\prime }|\mathbf{S}_{i}\cdot \mathbf{S}_{j}|h_{1},h_{2},\sigma
_{h}\rangle }{\langle \upsilon ^{\prime }|\upsilon \rangle }~.
\end{align}

%Distinguish different spin and loop configurations to categorize $E_{ij}^{J}\left( h,\upsilon ,\upsilon ^{\prime }\right) $

Given a transposition graph $\left( \upsilon ,\upsilon ^{\prime }\right) $,
we categorize $E_{ij}^{J}\left( h,\upsilon ,\upsilon ^{\prime }\right) $ in
a list below.

\begin{itemize}
\item $h_{1}$ and $h_{2}$ belong to the same loop $L_{h_{1}h_{2}}$ in a
given transposition graph~$\left( \upsilon ,\upsilon ^{\prime }\right) ~.$

\begin{enumerate}[ 1)]
\item One of the two holes coincides with the site $i$ or $j,$which gives $%
E_{ij}^{J}\left( h,\upsilon ,\upsilon ^{\prime }\right) =0~.$

\item Sites $i$ and $j$ belong to the different loops of the transposition
graph $\left( \upsilon ,\upsilon ^{\prime }\right) $. The contributions from
terms $S_{i}^{+}S_{j}^{-}+S_{i}^{-}S_{j}^{+}$ always vanish since a closed
loop can not have a single antiferromagnetic domain. Although the
expectation value of diagonal terms $S_{i}^{z}S_{j}^{z}$ for a fixed spin
configuration is not zero, their contributions vanish after summation of all
compatible spin configurations.

\item Sites $i,j$ belong to the same loop $L_{ij}$ that contains no holes, $%
L_{ij}\neq L_{h_{1}h_{2}}$. If the two holes belong to different sublattices 
$\delta _{h_{1}h_{2}}^{\text{sublatt}}=0$ (to satisfy compatibility), the
contribution $E_{ij}^{J}\left( h,\upsilon ,\upsilon ^{\prime }\right) $
reads 
\begin{align}
E_{ij}^{J}\left( h,\upsilon ,\upsilon ^{\prime }\right) & =2\times
2^{N_{\upsilon ,\upsilon ^{\prime }}^{\text{loop}}-2}\cdot \mathrm{Re}\left(
\Delta \Lambda _{ij}^{J}\right) \cdot \frac{-J/2}{2^{N_{\upsilon ,\upsilon
^{\prime }}^{\text{loop}}-2}}+2^{N_{\upsilon ,\upsilon ^{\prime }}^{\text{%
loop}}-1}\cdot 1\cdot \frac{-J/4}{2^{N_{\upsilon ,\upsilon ^{\prime }}^{%
\text{loop}}-2}}  \notag \\
& =2\left(-\frac{J}{2}\mathrm{Re}\left( \Delta \Lambda _{ij}^{J}\left( h\right)
\right) -\frac{J}{4}\right)~.
\end{align}%
Otherwise, $E_{ij}^{J}\left( h,\upsilon ,\upsilon ^{\prime }\right) =0.$

\item Sites $i,j$ belong to the loop $L_{h_{1}h_{2}}.$ This is more
complicated. If and only if holes $h_{1}$ and $h_{2}$ belong to different
sublattices, terms $S_{i}^{z}S_{j}^{z}$ contribute nonvanishingly to $%
E_{ij}^{J}\left( h,\upsilon ,\upsilon ^{\prime }\right) $. Terms $%
S_{i}^{+}S_{j}^{-}+S_{i}^{-}S_{j}^{+}$ depend on relative positions of sites 
$i,j$ and holes. For the sake of clarity, we introduce an auxiliary loop $%
L_{h_{1}h_{2}}^{\prime }$, which is obtained from the loop $L_{h_{1}h_{2}}$
by setting $\upsilon (\upsilon (i))=j,\upsilon (\upsilon (j))=i$ in loop $%
L_{h_{1}h_{2}}^{\prime }$. We also introduce an auxiliary dimmer
configuration $|h_{1},h_{2},\upsilon ^{\prime \prime }\rangle
=(S_{i}^{+}S_{j}^{-}+S_{i}^{-}S_{j}^{+})|h_{1},h_{2},\upsilon \rangle $, and
spin configuration \ $|h_{1},h_{2},\sigma _{h}^{\prime \prime }\rangle
=\left( S_{i}^{+}S_{j}^{-}+S_{i}^{-}S_{j}^{+}\right) |h_{1},h_{2},\sigma
_{h}\rangle $ ($\sigma _{h}^{\prime \prime }$ is compatible with
transposition graph $\left( \upsilon ^{\prime },\upsilon ^{\prime \prime
}\right) $). Note that the auxiliary loop $L_{h_{1}h_{2}}^{\prime }$ and
dimmer configuration $|h_{1},h_{2},\upsilon ^{\prime \prime }\rangle $\ do
not satisfy original rules of construction. If the spin configuration $%
\sigma _{h}^{\prime \prime }$ satisfies $\sigma _{h}^{\prime \prime
}(h_{1})=-\sigma _{h}^{\prime \prime }(h_{2})$, 
\begin{equation}
E_{ij}^{J}\left( h,\upsilon ,\upsilon ^{\prime }\right) =2\left[-\frac{J}{2}%
\mathrm{Re}\left( \Delta \Lambda _{ij}^{J}\left( h\right) \right) \delta
_{\sigma _{h}^{\prime \prime }(h_{1}),-\sigma _{h}^{\prime \prime }(h_{2})}-%
\frac{J}{4}\left(1-\delta _{h_{1}h_{2}}^{\text{sublatt}}\right)\right] ~,
\end{equation}%
where~$\left( 1-\delta _{h_{1}h_{2}}^{\text{sublatt}}\right) =1$ if the two
holes~$h_{1},h_{2}$~belong to the different sublattices.
\end{enumerate}

\item Holes $h_{1}~$and~$h_{2}$~belong to different loops $%
L_{h_{1}},L_{h_{2}}$ in a given transposition graph~$\left( \upsilon
,\upsilon ^{\prime }\right) $

\begin{enumerate}[ 1)]
\item If one of $h_{1}$ and $h_{2}$ coincides to site $i$ or $j,$ $%
E_{ij}^{J}\left( h,\upsilon ,\upsilon ^{\prime }\right) =0.$
\item Sites $i,j$ belong to different loops of $\left( \upsilon ,\upsilon
^{\prime }\right) $. Contributions from terms $%
S_{i}^{+}S_{j}^{-}+S_{i}^{-}S_{j}^{+}$ vanish for there exists no compatible
spin configuration with a VB state. Only when~$i\in L_{h_{1}},j\in L_{h_{2}}$
or $i\in L_{h_{2}},j\in L_{h_{1}}$, the diagonal terms $S_{i}^{z}S_{j}^{z}$
contributes to $E_{ij}^{J}\left( h,\upsilon ,\upsilon ^{\prime }\right) $,%
\begin{equation}
E_{ij}^{J}\left( h,\upsilon ,\upsilon ^{\prime }\right) =\frac{J}{4}\frac{%
2^{N_{\upsilon ,\upsilon ^{\prime }}^{\text{loop}}-2}}{2^{N_{\upsilon
,\upsilon ^{\prime }}^{\text{loop}}-2}}\left( \delta _{\sigma _{h}(i),\sigma
_{h}(j)}-\delta _{-\sigma _{h}(i),\sigma _{h}(j)}\right) =\frac{J}{4}\left(
\delta _{\sigma _{h}(i),\sigma _{h}(j)}-\delta _{-\sigma _{h}(i),\sigma
_{h}(j)}\right) ~.
\end{equation}

\item Site~$i,j$~belong to the same loop $L_{ij}$ of the transposition graph~%
$\left( \upsilon ,\upsilon ^{\prime }\right) $. If $L_{ij}$ does not contain
holes, we obtain%
\begin{align}
E_{ij}^{J}\left( h,\upsilon ,\upsilon ^{\prime }\right) & =2\times
2^{N_{\upsilon ,\upsilon ^{\prime }}^{\text{loop}}-3}\cdot \mathrm{Re}\left(
\Delta \Lambda _{ij}^{J}\right) \cdot \frac{-J/2}{2^{N_{\upsilon ,\upsilon
^{\prime }}^{\text{loop}}-2}}+2^{N_{\upsilon ,\upsilon ^{\prime }}^{\text{%
loop}}-2}\cdot 1\cdot \frac{-J/4}{2^{N_{\upsilon ,\upsilon ^{\prime }}^{%
\text{loop}}-2}}  \notag \\
& =-\frac{J}{2}\mathrm{Re}\left( \Delta \Lambda _{ij}^{J}\left( h\right)
\right) -\frac{J}{4}~.
\end{align}%
The case that $L_{ij}$ contains one or two holes shows the same result, but
only one term of ~$S_{i}^{+}S_{j}^{-}$~and~$S_{i}^{-}S_{j}^{+}$~helps.
\end{enumerate}
\end{itemize}

\subsection{Hopping energy}

\label{sec:Ht}

In this section, we turn to calculation of $\langle H^{t}\rangle $ 
\begin{equation}
\langle H^{t}\rangle =\frac{(\sum_{\upsilon ,\upsilon ^{\prime },\sigma
^{0}}\delta _{\upsilon ,\sigma ^{0}}\delta _{\upsilon ^{\prime },\sigma
^{0}})\omega _{\upsilon ^{\prime }}\omega _{\upsilon }E(\upsilon ,\upsilon
^{\prime })}{(\sum_{\upsilon ,\upsilon ^{\prime },\sigma ^{0}}\delta
_{\upsilon ,\sigma ^{0}}\delta _{\upsilon ^{\prime },\sigma ^{0}})}~,
\end{equation}%
where%
\begin{align}
E(\upsilon ,\upsilon ^{\prime })& =\sum_{\substack{ h_{1},h_{2}, \\ %
h_{1}\not=h_{2}}}\sum_{\substack{ h_{1}^{\prime },h_{2}^{\prime } \\ %
h_{1}^{\prime }\not=h_{2}^{\prime }}}E(h,h^{\prime },\upsilon ,\upsilon
^{\prime })~, \\
E(h,h^{\prime },\upsilon ,\upsilon ^{\prime })& =4\sum_{\sigma _{h^{\prime
}}^{\prime },\sigma _{h}^{{}}}\delta _{\upsilon ^{\prime },\sigma
_{h^{\prime }}^{\prime }}\delta _{\upsilon ,\sigma _{h}}^{{}}g^{\ast
}(h_{1}^{\prime },h_{2}^{\prime })g(h_{1},h_{2})\frac{\langle h_{1}^{\prime
},h_{2}^{\prime },\sigma _{h^{\prime }}^{\prime }|e^{i\hat{\Theta}}H^{t}e^{-i%
\hat{\Theta}}|h_{1},h_{2},\sigma _{h}\rangle }{\langle \upsilon ^{\prime
}|\upsilon \rangle }~.
\end{align}%
where we take $h$ $\left( h^{\prime }\right) $ in $E(h,h^{\prime },\upsilon
,\upsilon ^{\prime })$ as a shorthand for $h_{1}$ and $h_{2}$ $\left(
h_{1}^{\prime }\text{ and }h_{2}^{\prime }\right) $. Each hopping term only
moves one hole within a single action. Without loss of generality, we can
assume the position of hole $h_{1}$ unchanged, i.e. $h_{1}=h_{1}^{\prime }$.
Furthermore, similar to the trick upon terms $%
S_{i}^{+}S_{j}^{-}+S_{i}^{-}S_{j}^{+}$ in the superexchange energy, we
introduce an auxiliary spin and VB configuration:%
\begin{align}
|h_{1},h_{2}^{\prime },\sigma _{h}^{\prime \prime }\rangle &
=c_{h_{2}^{\prime }\uparrow }^{{}}c_{h_{2}\uparrow }^{\dagger
}|h_{1},h_{2},\sigma _{h}\rangle ~, \\
|h_{1},h_{2}^{\prime },\upsilon ^{\prime \prime }\rangle & =c_{h_{2}^{\prime
}\uparrow }^{{}}c_{h_{2}\uparrow }^{\dagger }|h_{1},h_{2},\upsilon \rangle ~.
\end{align}%
The auxiliary VB configuration $|h_{1},h_{2}^{\prime },\upsilon ^{\prime
\prime }\rangle $ requires $\upsilon ^{\prime \prime }\left( h_{2}^{\prime
}\right) =\upsilon \left( h_{2}\right) $ and other dimmers stay the same.
The non-vanishing contributions require compatibility between spin
configuration $|h_{1}^{\prime },h_{2}^{\prime },\sigma _{h}^{\prime \prime
}\rangle $ and a new transposition graph $\langle h_{1},h_{2}^{\prime
},\upsilon ^{\prime }|h_{1},h_{2}^{\prime },\upsilon ^{\prime \prime
}\rangle .$The expression of $E(h,h^{\prime },\upsilon ,\upsilon ^{\prime })$
can be decomposed into several factors 
\begin{equation}
E(h,h^{\prime },\upsilon ,\upsilon ^{\prime })=-\sum_{\sigma _{h},\sigma
_{h}^{\prime }}\Delta \left( \sigma _{h},\sigma _{h}^{\prime }\right) \cdot
g^{\ast }\left( h_{1}^{\prime },h_{2}^{\prime }\right) g\left(
h_{1},h_{2}\right) \cdot \frac{4}{2^{n}}\cdot \Delta \Lambda ~,
\end{equation}%
where $n$ is the number of loops including sites $h_{1},h_{2}$ and $%
h_{2}^{\prime }$ in $\left\langle \upsilon ^{\prime }|\upsilon \right\rangle
,$ and $\Delta \left( \sigma _{h},\sigma _{h}^{\prime }\right) $ is the
Marshall sign difference between the initial $\sigma _{h}^{{}}$ and final
states $\sigma _{h}^{\prime }$. The minus sign comes from the permutation of
Fermions. The phase difference $\Delta \Lambda ,$ induced by phase string
effect, can be divided into four parts:%
\begin{equation}
\Delta \Lambda =\Delta \Lambda _{0}\cdot \Delta \Lambda _{1}\cdot \Delta
\Lambda _{2}\cdot \Delta \Lambda _{3}~.
\end{equation}

\begin{itemize}
\item $\Delta \Lambda _{0}$ comes from sites $h_{1},h_{2}$ and $%
h_{2}^{\prime }$:%
\begin{equation}
\Delta \Lambda _{0}=\Lambda ^{\ast }\left( h_{2}^{\prime },h_{2}^{{}},\sigma
_{h}\left( h_{2}^{\prime }\right) \right) \Lambda (h_{2}^{{}},h_{2}^{\prime
},\sigma _{h}\left( h_{2}^{\prime }\right) )\Lambda ^{\ast }\left(
h_{1}^{{}},h_{2}^{{}},\sigma _{h}\left( h_{2}^{\prime }\right) \right)
\Lambda \left( h_{1}^{{}},h_{2}^{\prime },\sigma _{h}\left( h_{2}^{\prime
}\right) \right) ~,
\end{equation}%
with $\Lambda \left( h,l,\sigma _{h}\left( l\right) \right) $ defined in (%
\ref{phasefactorOfaSite}).

\item $\Delta \Lambda _{1}$ comes from sites in the VB configuration $%
\langle h_{1},h_{2}^{\prime },\upsilon ^{\prime }|h_{1},h_{2}^{\prime
},\upsilon ^{\prime \prime }\rangle $ loops $L$ that contain sites $%
h_{1}^{\prime }\left( =h_{1}\right) $ or $h_{2}^{\prime }$ but except sites
that coincide with $h_{1}\left( =h_{1}^{\prime }\right) ,h_{2}$ or $%
h_{2}^{\prime }$. 
\begin{equation}
\Delta \Lambda _{1}=\prod_{_{\substack{ l\in L_{h_{1}^{\prime
}},L_{h_{2}^{\prime }}  \\ l\not=h_{1},h_{2},h_{1}^{\prime },h_{2}^{\prime }
}}}\Lambda ^{\ast }\left( h_{2}^{\prime },l,\sigma _{h}\left( l\right)
\right) \Lambda \left( h_{2},l,\sigma _{h}\left( l\right) \right) ~,
\end{equation}%
where $L_{h_{1}^{\prime }}$ ($L_{h_{2}^{\prime }}$) is the loop containing $%
h_{1}^{\prime }$ ($h_{2}^{\prime }$).

\item If neither of the loops $L_{h_{1}}$ nor $L_{h_{2}^{\prime }}$ of the
VB configuration\ $\langle h_{1},h_{2}^{\prime },\upsilon ^{\prime
}|h_{1},h_{2}^{\prime },\upsilon ^{\prime \prime }\rangle $ contain the site 
$h_{2}$, that is $L_{h_{2}}\neq L_{h_{1}}$ and $L_{h_{2}}\neq
L_{h_{2}^{\prime }}$, there are two different spin configurations that are
compatible with the loop $L_{h_{2}}$, which account for the phase factor $%
\Delta \Lambda _{2}$. 
\begin{equation}
\Delta \Lambda _{2}=\sum_{\sigma _{l}=\pm }\prod_{l\in L_{h_{1}}}\Lambda
^{\ast }\left( h_{2}^{\prime },l,\sigma _{h}\left( l\right) \right) \Lambda
\left( h_{2},l,\sigma _{h}\left( l\right) \right) ~.
\end{equation}%
Otherwise, $\Delta \Lambda _{2}=1$.

\item $\Delta \Lambda _{3}$ comes from the rest loops of VB configuration $%
\langle h_{1},h_{2}^{\prime },\upsilon ^{\prime }|h_{1},h_{2}^{\prime
},\upsilon ^{\prime \prime }\rangle $%
\begin{equation}
\Delta \Lambda _{3}=\prod_{L\neq L_{h_{1}},L_{h_{2}},L_{h_{2}^{\prime }}} 
\left[ \frac{1}{2}\sum_{\left\{ \sigma _{l}\right\} }\prod_{l\in L}\Lambda
^{\ast }\left( h_{2}^{\prime },l,\sigma _{h}\left( l\right) \right) \Lambda
\left( h_{2},l,\sigma _{h}\left( l\right) \right) \right] ~.
\end{equation}
\end{itemize}

\subsection{Pair-pair correlation}

One may examine the pair-pair correlators, $\left\langle \Delta
_{ij}^{s,t}\left( \Delta _{ij}^{s,t}\right) ^{\dagger }\right\rangle $ where
the singlet/triplet channels are defined as follows%
\begin{align}
\Delta _{ij}^{s}& =\frac{1}{\sqrt{2}}\sum_{\sigma }\sigma c_{1i\sigma
}c_{2j-\sigma }~, \\
\Delta _{ij}^{t}& =\frac{1}{\sqrt{2}}\sum_{\sigma }c_{1i\sigma }c_{2j-\sigma
}~.
\end{align}%
Expand the correlators, 
\begin{equation}
C(i,j)=\left\langle \Delta _{ij}^{s,t}\left( \Delta _{ij}^{s,t}\right)
^{\dagger }\right\rangle =\left\langle c_{1i\downarrow }^{{}}c_{2j\uparrow
}^{{}}c_{2j\uparrow }^{\dagger }c_{1i\downarrow }^{\dagger }\mp
c_{1i\uparrow \downarrow }^{{}}c_{2j\downarrow }^{{}}c_{2j\uparrow
}^{\dagger }c_{1i\downarrow }^{\dagger }\right\rangle ~.
\end{equation}%
For the simulation of the pair-pair correlators, we only have to deal with
terms like%
\begin{equation}
C\left( h^{\prime },h\right) =\left\langle c_{h_{1}^{\prime }\downarrow
}c_{h_{2}^{\prime }\uparrow }c_{h_{2}\uparrow }^{\dagger }c_{h_{1}\downarrow
}^{\dagger }\right\rangle ~.
\end{equation}%
Here $h_{1}^{\prime }$ and $h_{2}^{\prime }$ correspond to the hole $h_{1}$
and $h_{2}$ with the same spin index respectively. Some simple operations
give%
\begin{equation}
C\left( h^{\prime },h\right) =\frac{(\sum_{\upsilon ,\upsilon ^{\prime
},\sigma ^{0}}\delta _{\upsilon ,\sigma ^{0}}\delta _{\upsilon ^{\prime
},\sigma ^{0}})\omega _{\upsilon ^{\prime }}\omega _{\upsilon }\mathcal{C}%
_{h^{\prime }h}}{(\sum_{\upsilon ,\upsilon ^{\prime },\sigma ^{0}}\delta
_{\upsilon ,\sigma ^{0}}\delta _{\upsilon ^{\prime },\sigma ^{0}})\omega
_{\upsilon ^{\prime }}\omega _{\upsilon }}~,
\end{equation}%
where%
\begin{equation}
\mathcal{C}_{h^{\prime }h}=\frac{4\left\langle h_{1}^{\prime },h_{2}^{\prime
},\upsilon ^{\prime }\left\vert e^{i\Theta }c_{h_{1}^{\prime }\downarrow
}c_{h_{2}^{\prime }\uparrow }c_{h_{2}\uparrow }^{\dagger }c_{h_{1}\downarrow
}^{\dagger }e^{-i\Theta }\right\vert h_{1},h_{2},\upsilon \right\rangle
g^{\ast }\left( h_{1}^{\prime },h_{2}^{\prime }\right) g\left(
h_{1},h_{2}\right) }{\left\langle \upsilon ^{\prime }|\upsilon \right\rangle 
}~.
\end{equation}%
Introduce an auxiliary spin and VB configuration:%
\begin{align}
|h_{1}^{\prime },h_{2}^{\prime },\sigma _{h}^{\prime \prime }\rangle &
=c_{h_{1}^{\prime }\downarrow }^{{}}c_{h_{2}^{\prime }\uparrow
}^{{}}c_{h_{2}\uparrow }^{\dagger }c_{h_{1}\downarrow }^{\dagger
}|h_{1}^{{}},h_{2}^{{}},\sigma _{h}^{{}}\rangle \\
|h_{1}^{\prime },h_{2}^{\prime },\upsilon ^{\prime \prime }\rangle &
=c_{h_{1}^{\prime }\downarrow }^{{}}c_{h_{2}^{\prime }\uparrow
}^{{}}c_{h_{2}\uparrow }^{\dagger }c_{h_{1}\downarrow }^{\dagger
}|h_{1}^{{}},h_{2}^{{}},\upsilon \rangle ~.
\end{align}%
The auxiliary VB configuration $|h_{1}^{\prime },h_{2}^{\prime },\upsilon
^{\prime \prime }\rangle $ requires $\upsilon ^{\prime \prime }\left(
h_{1}^{\prime }\right) =\upsilon \left( h_{1}\right) $,$\upsilon ^{\prime
\prime }\left( h_{2}^{\prime }\right) =\upsilon \left( h_{2}\right) $ and
other dimmers stay the same. The nonvanishing contributions require
compatibility of spin configuration $|h_{1}^{\prime },h_{2}^{\prime },\sigma
_{h}^{\prime \prime }\rangle $ with a new transposition graph $\langle
h_{1}^{\prime },h_{2}^{\prime },\upsilon ^{\prime }|h_{1}^{\prime
},h_{2}^{\prime },\upsilon ^{\prime \prime }\rangle .$

The factor $C$ can be decomposed into several parts:%
\begin{equation}
\mathcal{C}_{h^{\prime }h}=\sum_{\sigma _{h},\sigma _{h}^{\prime }}\Delta
\left( \sigma _{h},\sigma _{h}^{\prime }\right) \cdot g^{\ast }\left(
h_{1}^{\prime },h_{2}^{\prime }\right) g\left( h_{1},h_{2}\right) \cdot 
\frac{4}{2^{n}}\cdot \Delta \Lambda ~,
\end{equation}%
where 
\begin{align}
\Delta \left( \sigma _{h},\sigma _{h}^{\prime }\right) & \text{:}\text{
Marshall sign difference between initial }\sigma _{h}\text{ and final states 
}\sigma _{h}^{\prime } \\
n& \text{: the number of loops including sites }h_{1},h_{2},h_{1}^{\prime }%
\text{ and }h_{2}^{\prime }\text{ in }\left\langle \upsilon ^{\prime
}|\upsilon \right\rangle \text{ }~.
\end{align}%
The phase difference $\Delta \Lambda $ induced by phase string effect can be
divided into four parts:%
\begin{equation}
\Delta \Lambda =\Delta \Lambda _{0}\cdot \Delta \Lambda _{1}\cdot \Delta
\Lambda _{2}\cdot \Delta \Lambda _{3}~.
\end{equation}

\begin{itemize}
\item $\Delta \Lambda _{0}$ comes from sites $h_{1},h_{2},h_{1}^{\prime }$
and $h_{2}^{\prime }$:%
\begin{equation}
\Delta \Lambda _{0}=\prod_{l=1,2}\Lambda ^{\ast }\left( h_{l}^{\prime
},h_{2},\sigma _{l}\right) \Lambda ^{\ast }\left( h_{l}^{\prime
},h_{2},\sigma _{l}\right) \Lambda \left( h_{l},h_{1}^{\prime },\sigma
_{l}\right) \Lambda \left( h_{l},h_{2}^{\prime },\sigma _{l}\right)~,
\end{equation}%
where $\sigma _{1}=\sigma _{h}\left( h_{1}^{\prime }\right) $ and $\sigma
_{2}=\sigma _{h}\left( h_{2}^{\prime }\right) $.

\item $\Delta \Lambda _{1}$ comes from sites in the VB configuration $%
\left\langle \upsilon ^{\prime }|\upsilon ^{\prime \prime }\right\rangle $
loops that contains sites $h_{1}^{\prime }$ and $h_{2}^{\prime }$, except
sites $h_{1},h_{2},h_{1}^{\prime },h_{2}^{\prime }$. 
\begin{equation}
\Delta \Lambda _{1}=\prod_{\substack{ l\in L_{h_{1}^{\prime
}},L_{h_{2}^{\prime }}  \\ l\neq h_{1},h_{2},h_{1}^{\prime },h_{2}^{\prime } 
}}\Lambda ^{\ast }\left( h_{1}^{\prime },l,\sigma _{l}\right) \Lambda ^{\ast
}\left( h_{2}^{\prime },l,\sigma _{l}\right) \Lambda \left( h_{1},l,\sigma
_{l}\right) \Lambda \left( h_{2},l,\sigma _{l}\right) ~.
\end{equation}%
where $\sigma _{l}=\sigma _{h}\left( l\right) .$

\item Similar to the discussion in Sec \ref{sec:Ht}, we list cases for $%
\Delta \Lambda _{2}$.

\begin{enumerate}[ 1)]

\item $h_{1}\notin L_{h_{1}^{\prime }}\cup L_{h_{2}^{\prime }}$ and $%
h_{2}\in L_{h_{1}}$ in $\left\langle \upsilon ^{\prime }|\upsilon ^{\prime
\prime }\right\rangle $; Or $h_{1}\notin L_{h_{1}^{\prime }}\cup
L_{h_{2}^{\prime }}$ and $h_{2}\notin L_{h_{1}},h_{2}\in L_{h_{1}^{\prime
}}\cup L_{h_{2}^{\prime }}$ in $\langle h_{1}^{\prime },h_{2}^{\prime
},\upsilon ^{\prime }|h_{1}^{\prime },h_{2}^{\prime },\upsilon ^{\prime
\prime }\rangle $ 
\begin{equation}
\Delta \Lambda _{2}=\sum_{\sigma _{l}}\prod_{\substack{ l\in L_{h_{1}}  \\ %
l\neq h_{1},h_{2}}}\Lambda ^{\ast }\left( h_{1}^{\prime },l,\sigma
_{l}\right) \Lambda ^{\ast }\left( h_{2}^{\prime },l,\sigma _{l}\right)
\Lambda \left( h_{1},l,\sigma _{l}\right) \Lambda \left( h_{2},l,\sigma
_{l}\right) ~;
\end{equation}

\item $h_{2}\notin L_{h_{1}^{\prime }}\cup L_{h_{2}^{\prime }}$ and $%
h_{1}\notin L_{h_{2}}$ in $\langle h_{1}^{\prime },h_{2}^{\prime },\upsilon
^{\prime }|h_{1}^{\prime },h_{2}^{\prime },\upsilon ^{\prime \prime }\rangle 
$%
\begin{equation}
\Delta \Lambda _{2}=\sum_{\sigma _{l}}\prod_{\substack{ l\in L_{h_{2}}  \\ %
l\neq h_{2}}}\Lambda ^{\ast }\left( h_{1}^{\prime },l,\sigma _{l}\right)
\Lambda ^{\ast }\left( h_{2}^{\prime },l,\sigma _{l}\right) \Lambda \left(
h_{1},l,\sigma _{l}\right) \Lambda \left( h_{2},l,\sigma _{l}\right) ~;
\end{equation}

\item $h_{1}\notin L_{h_{1}^{\prime }}\cup L_{h_{2}^{\prime }},h_{2}\notin
L_{h_{1}^{\prime }}\cup L_{h_{2}^{\prime }}$ and $L_{h_{1}}\neq L_{h_{2}}$
in $\langle h_{1}^{\prime },h_{2}^{\prime },\upsilon ^{\prime
}|h_{1}^{\prime },h_{2}^{\prime },\upsilon ^{\prime \prime }\rangle $%
\begin{equation}
\Delta \Lambda _{2}=\prod_{L=L_{h_{1}},L_{h_{2}}}\sum_{\sigma _{l}}\prod 
_{\substack{ l\in L  \\ l\neq h_{1},h_{1}}}\Lambda ^{\ast }\left(
h_{1}^{\prime },l,\sigma _{l}\right) \Lambda ^{\ast }\left( h_{2}^{\prime
},l,\sigma _{l}\right) \Lambda \left( h_{1},l,\sigma _{l}\right) \Lambda
\left( h_{2},l,\sigma _{l}\right) ~;
\end{equation}

\item Otherwise%
\begin{equation}
\Delta \Lambda _{2}=1~.
\end{equation}
\end{enumerate}

Here, the notation $L_{h_{1}^{\prime }}\cup L_{h_{2}^{\prime }}$ represents
the set containing all sites from $L_{h_{1}^{\prime }}$ and $%
L_{h_{2}^{\prime }}$ in $\left\langle \upsilon ^{\prime }|\upsilon ^{\prime
\prime }\right\rangle $.

\item $\Delta \Lambda _{3}$ comes from the remaining parts of the VB configuration $%
\langle h_{1}^{\prime },h_{2}^{\prime },\upsilon ^{\prime }|h_{1}^{\prime
},h_{2}^{\prime },\upsilon ^{\prime \prime }\rangle $%
\begin{equation}
\Delta \Lambda _{3}=\prod_{L\not=L_{h_{1}},L_{h_{2}},L_{h_{1}^{\prime
}},L_{h_{2}^{\prime }}}\left[ \frac{1}{2}\sum_{\left\{ \sigma _{l}\right\}
}\prod_{l\in L}\Lambda ^{\ast }\left( h_{1}^{\prime },l,\sigma _{l}\right)
\Lambda ^{\ast }\left( h_{2}^{\prime },l,\sigma _{l}\right) \Lambda \left(
h_{1},l,\sigma _{l}\right) \Lambda \left( h_{2},l,\sigma _{l}\right) \right]
~,
\end{equation}%
where $\sigma _{l}=\sigma _{h}\left( l\right) $.
\end{itemize}

\section{DMRG results of the two-hole doped $t$-$J$ two-leg ladder with $t_\perp>0$ } \label{appden:ty>0}
We investigate two-hole pairing in the limit of $t_\perp=0$ for the two-leg ladder in the main body of this work. In order to show that the non-BCS pairing discovered there can be qualitatively applied to a more general case, here we present the numerical results of the two-hole ground state from  $t_\perp=0$ to $t_\perp>0$ with  $\alpha=1$ by DMRG.  Indeed, a smooth crossover without any ``phase transition'' is shown by the first and second derivatives of the ground state energy versus $t_\perp$ over a finite range of $t_\perp\geq 0$ as illustrated by Fig. \ref{pair_cor}.
\begin{figure}[ht]
\begin{center}
\includegraphics[width=1\textwidth]{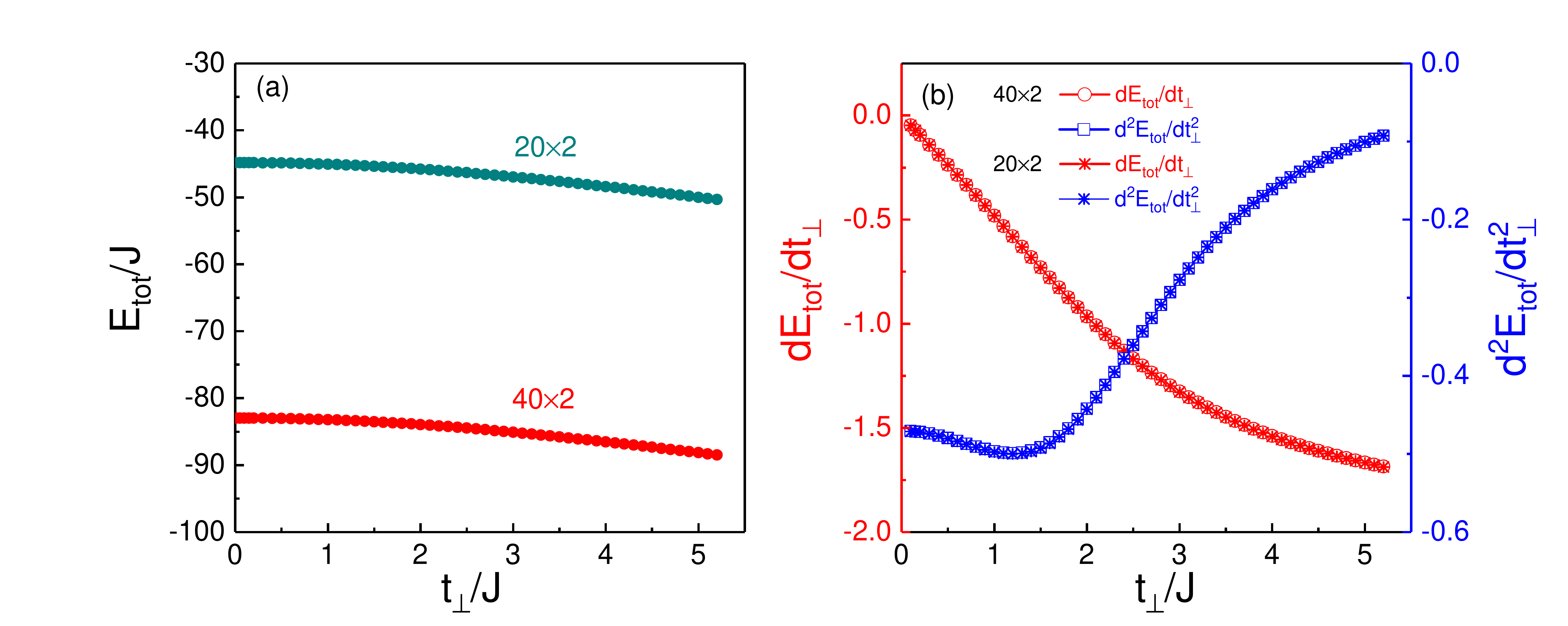}
\end{center}
\par
\renewcommand{\figurename}{Fig.}
\caption{(Color online.) (a) The total energy and (b), the first and second derivatives with respect to $t_\perp$ with $\alpha=1$.  We calculate two different lattice sizes $20\times 2$ and $40\times2$. }
\label{Fig:Energy-ty}
\end{figure}
\newpage 
%\bibliographystyle{apsrev4-1}
%\bibliography{refs}
%%merlin.mbs apsrev4-1.bst 2010-07-25 4.21a (PWD, AO, DPC) hacked
%Control: key (0)
%Control: author (72) initials jnrlst
%Control: editor formatted (1) identically to author
%Control: production of article title (-1) disabled
%Control: page (0) single
%Control: year (1) truncated
%Control: production of eprint (0) enabled

%merlin.mbs apsrev4-1.bst 2010-07-25 4.21a (PWD, AO, DPC) hacked
%Control: key (0)
%Control: author (72) initials jnrlst
%Control: editor formatted (1) identically to author
%Control: production of article title (-1) disabled
%Control: page (0) single
%Control: year (1) truncated
%Control: production of eprint (0) enabled
%

\end{document}